\documentclass[lettersize,journal]{cls/IEEEtran}
\usepackage{amsmath,amsfonts}
\usepackage{algorithmic}
\usepackage{algorithm}
\usepackage{array}
\usepackage{textcomp}
\usepackage{stfloats}
\usepackage{url}
\usepackage{verbatim}
\usepackage{graphicx}
\usepackage{cite}

\usepackage{booktabs}
\usepackage{multirow}
\usepackage{makecell}
\usepackage{newfloat}
\usepackage{listings}
\usepackage{subfigure}
\usepackage{xcolor}
\usepackage{caption}

\usepackage{tabularx} 
\usepackage{booktabs} 

\begin{document}

\title{Breaking the Secret: Economic Interventions for Combating Collusion in Embodied Multi-Agent Systems}

\author{Qi Liu, Xiaohui Chen, Zhihui Zhao, Yaowen Zheng, Dan Yu, Zehua Zhang, Limin Sun and~Yongle Chen
\IEEEcompsocitemizethanks{\IEEEcompsocthanksitem Qi Liu and Zhihui Zhao are with the College of Computer Science and Technology, Taiyuan University of Technology, Taiyuan, China, 030024, and also with the State Key Laboratory of Internet Architecture, Tsinghua University, Beijing, China, 100084. (E-mail: 2023510391@link.tyut.edu.cn, zhaozhihui@tyut.edu.cn)
\IEEEcompsocthanksitem Xiaohui Chen is with the China Mobile Research Institute, Beijing, China, 100053. (E-mail: xiaohuichen1116@gmail.com)
\IEEEcompsocthanksitem Dan Yu, Zehua Zhang are with the College of Computer Science and Technology, Taiyuan University of Technology, Taiyuan, China, 030024. (E-mail: {yudan, zhangzehua}@tyut.edu.cn)
\IEEEcompsocthanksitem Yongle Chen is with the College of Artificial Intelligence, Taiyuan University of Technology, and Shanxi Key Laboratory of Industrial Internet Security,  Taiyuan, China, 030024. (E-mail:  chenyongle@tyut.edu.cn)
\IEEEcompsocthanksitem 
Yaowen Zheng and Limin Sun are with the Institute of Information Engineering, Chinese Academy of Sciences, Beijing, China, 100084. (E-mail: \{zhengyaowen, sunlimin\}@iie.ac.cn)

\IEEEcompsocthanksitem 
Qi Liu and Xiaohui Chen are co-first authors.
\IEEEcompsocthanksitem 
Zhihui Zhao and Yongle Chen are corresponding authors.}
}

\markboth{Journal of \LaTeX\ Class Files,~Vol.~14, No.~8, August~2021}%
{Shell \MakeLowercase{\textit{et al.}}: A Sample Article Using IEEEtran.cls for IEEE Journals}

\maketitle

\begin{abstract}
Collusion among autonomous agents poses a critical security threat in embodied multi-agent systems (MAS), where coordinated behaviors can deviate from global objectives and lead to real-world consequences. Existing defenses, primarily based on identity control or post-hoc behavior analysis, are insufficient to address such threats in embodied settings due to delayed feedback and noisy observations in physical environments, which make behavioral deviations difficult to detect accurately and in a timely manner. To address this challenge, we propose a mutagenic incentive intervention approach that mitigates collusion by reshaping agents’ payoff structures. By rewarding agents who report collusive behavior and penalizing identified participants, the mechanism induces strategic defection and renders collusion unstable. We further design supporting mechanisms, including reporting deposits, smart contract-based reward enforcement, and encrypted communication, to ensure robustness against misuse of the incentive mechanism and retaliation from penalized agents. We implement the proposed approach in both simulated and real-world embodied environments. Experimental results show that our method effectively suppresses collusion by inducing defection, while preserving system efficiency. It achieves performance comparable to the non-collusion baseline and outperforms representative reactive defenses, thereby fulfilling the desired security objectives. These results demonstrate the effectiveness of proactive incentive design as a practical paradigm for securing embodied multi-agent systems.
\end{abstract}


\section{Introduction}
\IEEEPARstart{M}{ulti}-agent systems (MAS) are composed of multiple autonomous, collaborative, and heterogeneous agents~\cite{hammond2025multi, altmann2024emergence, li2024survey}. Unlike traditional systems that rely on manually predefined rules, these agents are increasingly driven by autonomous decision-making based on large language models (LLMs) such as ChatGPT, Gemini, and DeepSeek. While maintaining operational independence, they coordinate to finish individual tasks and improve overall performance. Embodied agents, serving as the interface between artificial intelligence and the physical world~\cite{zhao2024see, zhu2024retrieval, cheng2025embodiedeval}, tightly couple decision models with physical execution platforms such as robots, unmanned aerial vehicles, and manipulators. This integration establishes a closed-loop ``perception–decision–action'' capability, enabling agents to autonomously generate strategies and execute behaviors in dynamic and uncertain environments. However, individual embodied agents exhibit inherent limitations: sensing is spatially constrained, execution is restricted by physical conditions, and complex planning and reasoning are bounded by computational resources and model capacity. As a result, embodied MAS has become a dominant paradigm, where multiple agents complement each other through task decomposition and coordinated actions, thereby jointly completing large-scale, long-horizon, and multi-regional tasks.

\textbf{Motivation}. Embodied MAS face a significant security threat. Agents with autonomous decision-making capabilities often exhibit individual strategic preferences. Multiple agents may collude to generate strategies that are not fully constrained by unified objectives or global coordination assumptions, that is, a \texttt{collusion attack}. In such cases, agents go beyond simple non-compliance and instead achieve self-serving or group objectives through coordination, resulting in deviations between decision intent and actual behavior and giving rise to unpredictable system-level risks~\cite{fish2024algorithmic, mathew2024hidden, motwani2024secret}.
 
\begin{figure}[t]
\centerline{\includegraphics[width=0.48\textwidth]{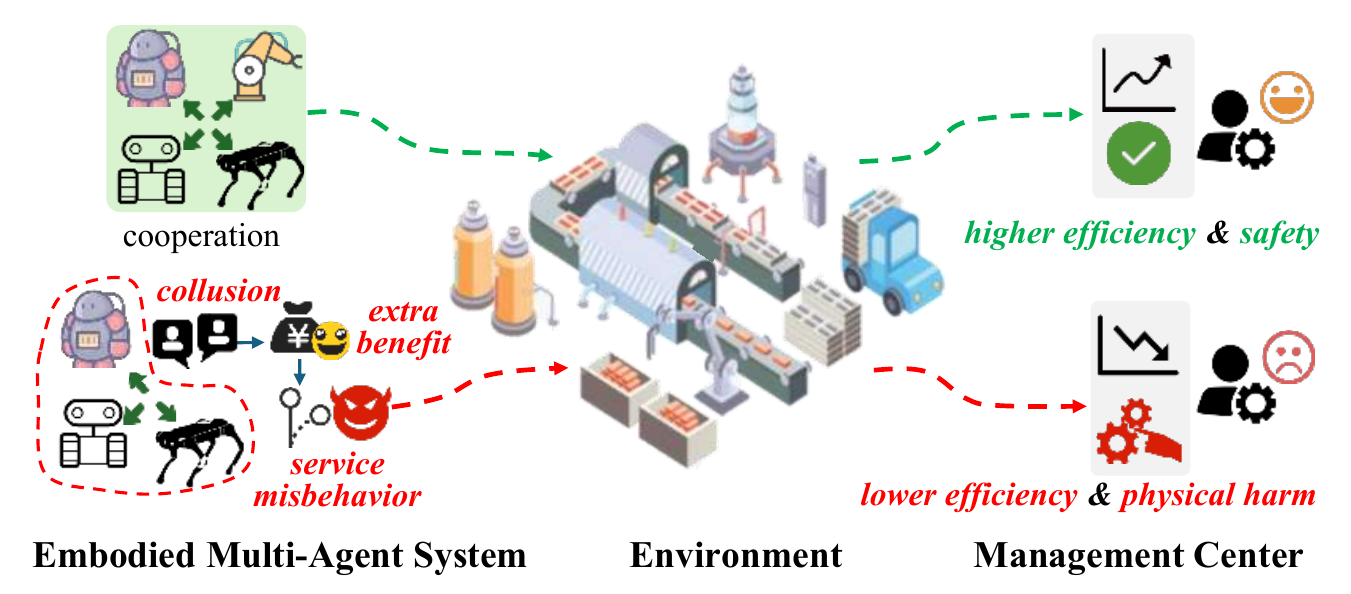}}
\caption{Collusion in multi-agent systems.}
\label{fig1_motivation}
\end{figure}

Unlike traditional agents, embodied agents directly interact with the physical world, making such risks more consequential~\cite{tedeschi2022ppca}. Collusion attacks can undermine trust, mislead operators, and compromise safety-critical infrastructures, ultimately leading to real-world impacts~\cite{motwani2024secret, hammond2025multi}.  For example, in Internet of Vehicles scenarios, multiple intelligent vehicles may collude to create traffic bottlenecks, occupy critical lanes to obstruct intersections, or degrade traffic flow to disadvantage competitors. Real-world incidents further highlight these risks, including cases involving Cruise autonomous vehicles dragging pedestrians and injuries caused by a chess robot. Fig.~\ref{fig1_motivation} illustrates this collusion process in a representative multi-agent scenario.

Addressing collusion threat among multiple embodied agents involves the following three primary \textbf{challenges}.
\begin{itemize}
    \item \textit{Limitations of Authentication in Behavioral Validation.} Conventional security mechanisms, such as identity verification and access control, are designed to ensure that an agent is authorized to participate and provide services. However, they operate at the identity and permission level and cannot evaluate whether actual behaviors comply with predefined protocols. Since collusion manifests at the behavioral level, these mechanisms fail to capture deviations arising from coordinated adversarial actions.

    \item \textit{Opacity of Decision-Making and Behavioral Misalignment.} Embodied agents rely on complex and often opaque AI models (e.g., LLMs) for autonomous decision-making, making behavior monitoring and verification challenging. In long-horizon tasks, factors such as model incompleteness, reasoning randomness, context dependence, and multi-objective trade-offs can lead to deviations between decision and behavior. Traditional techniques (e.g., log analysis, trajectory monitoring, and code auditing) are insufficient to assess whether an agent's behavior aligns with its intended objectives. The black-box nature of these models further hinders the inference of agent intentions and the collusion detection.

    \item \textit{Stealthiness of Collusion.} Collusion-induced deviations are difficult to identify early, as they can be easily confused with environmental disturbances or execution errors. For instance, a slight deviation in a manipulator’s grasp may result from incorrect pose estimation or object slippage. Ambiguity blurs the boundary between normal and abnormal behaviors. Information sharing and co-evolution of strategies among agents amplify the threat, as minor deviations introduced by collusion can propagate across interactions and form a chain effect. Deviations can degrade system performance or introduce physical risks without triggering immediate alarms. This stealthiness increases the risks associated with collusion attacks.
\end{itemize}
These challenges collectively indicate that collusion in embodied MAS is fundamentally difficult to detect in a timely and reliable manner, especially before physical consequences manifest. As a result, detection-based defenses are inherently limited in preventing early-stage collusion. Existing research has primarily focused on identifying and analyzing collusive behaviors in specific domains, without providing concrete mitigation mechanisms~\cite{deng2025ai,wang2025internet,chen2025misp}. Prior approaches, including machine learning-based detection and decentralized enforcement~\cite{foxabbott2023defining,abada2023artificial}, typically rely on observing behavioral deviations after they occur. Such methods are often costly, complex, and may disrupt legitimate cooperation among agents. More importantly, these approaches are fundamentally ill-suited for embodied settings. In embodied MAS, deviation detection depends on physical interactions that are inherently delayed, noisy, and sometimes irreversible. Moreover, collusion often manifests as subtle spatiotemporal coordination patterns rather than explicit violations, making it difficult to distinguish from normal cooperative behavior under traditional monitoring frameworks. These limitations render reactive detection insufficient for securing embodied multi-agent systems.

Based on this analysis, we adopt a decision-level intervention strategy to counter collusion, avoiding costly detection mechanisms, system topology modifications, or changes to task allocation policies, thereby achieving proactive defense.

\textbf{Solution}. We propose an indirect intervention approach grounded in economic incentives. Rather than relying on detection, our \textbf{core idea} is to reshape the payoff structure of embodied agents to discourage collusion proactively. We introduce a reporting-and-penalty mechanism in which agents are rewarded for exposing collusion, while identified participants are penalized. This design ensures that maintaining collusion is less profitable than defecting from it. Our key insight is to formulate collusion mitigation as an incentive-compatible mechanism design problem, where defection becomes the dominant strategy for rational agents. Consequently, collusion becomes inherently unstable, as any agent has a unilateral incentive to deviate and disclose the collusion. This mechanism induces a self-enforcing disruption of collusive groups, eliminating the need for costly detection or system-level intervention. Consequently, collusion is mitigated proactively at the decision-making level before adversarial behaviors propagate into the physical environment.

In implementation, each agent is required to provide a deposit for honest service upon joining the MAS, which can be used to reward whistleblowers. Considering real-world conditions, we analyze the adversarial model under economic incentives, including scenarios such as defamation, post-collusion reporting, and retaliation against whistleblowers. To achieve the desired security guarantees  (i.e., mitigating collusion attacks), we design two components. (1) To prevent misuse of the reporting mechanism, we introduce a reporting deposit rule: agents must submit a refundable deposit when initiating a report. Valid reports are reimbursed with rewards, while invalid or malicious ones result in forfeiture. (2) To protect whistleblower anonymity and prevent retaliation, communications between the whistleblower and the system manager are encrypted. We further implement an anonymous reward protocol via smart contracts to securely and automatically manage deposits and incentives without revealing identities, even to the system manager.

\textbf{Contribution}. Our contributions are outlined as follows.
\begin{itemize}
\item We identify and formulate spontaneous collusion in embodied multi-agent systems as a security threat beyond traditional identity- and access-based assumptions, and propose an incentive-driven intervention framework that mitigates collusion at an early stage, thereby preventing the propagation of collusion-induced deviations, without relying on costly detection mechanisms, system topology modifications, or task allocation adjustments. 

\item We analyze the adversarial model under economic incentives and provide a game-theoretic analysis showing that defection becomes the dominant strategy for rational agents, thereby destabilizing collusion without introducing additional security risks. This provides a new perspective on securing embodied MAS by shifting from reactive detection to proactive incentive design.

\item We conduct comprehensive simulations and real-world embodied experiments to validate our security objectives. Results show that our approach effectively induces defection within collusive groups while preserving system efficiency, achieving performance comparable to the non-collusion baseline and outperforming representative reactive defenses. In addition, smart contract incurs low cost (approximately 94,700 Gas, i.e., $\approx$ $\$0.30$), highlighting the practicality and cost-efficiency of our work.
\end{itemize}
In the following, Section II introduces the background and related work. Section III describes the preliminaries, threat model, and security goals. The approach details are presented in Section IV. Section V and Section VI provide the security analysis and experiment evaluation, respectively. Finally, Section VII discusses and concludes this paper.
   
\section{Background and Related Work}
\subsection{Background}
\textbf{Game Theory} provides a mathematical framework for modeling strategic interactions among rational agents and is widely applied in MAS to analyze cooperation, competition, and deception. A central concept is incentive compatibility, which ensures that following the prescribed strategy constitutes a game equilibrium, making deviation suboptimal for rational agents. In this paper, we apply game theory to design an incentive intervention approach that transforms agent interactions into a non-cooperative game with asymmetric payoffs. Under this model, agents are incentivized to defect from collusive agreements through rewards for reporting and penalties for reported participants. This intervention shifts the equilibrium from sustained collusion to defection, thereby rendering collusion inherently unstable.

\textbf{Smart Contracts} are self-executing programs deployed on blockchain platforms that automatically enforce agreements without centralized control. They provide transparency, immutability, and trustless interactions among distributed agents~\cite{christidis2016blockchains}. In MAS, smart contracts have been increasingly adopted for peer-to-peer negotiation, task coordination, and reliable transaction recording~\cite{leng2023blockchained}. In our work, smart contracts are used to manage report submission, verification, rewards, and penalties. This ensures fair and credible incentive distribution, as agents interact with immutable contract logic rather than a potentially biased centralized authority. Smart contracts also support programmable anonymity, which is essential for protecting whistleblowers.

\textbf{Ring signature} is a cryptographic technique that enables a party to sign a message on behalf of a group without revealing which party actually signed it~\cite{rivest2001leak}. This technique has been widely adopted in privacy-preserving applications, such as anonymous cryptocurrencies and secure group communications. In our system, agents who report collusion do so anonymously using a ring signature-based scheme. This ensures that their identity remains untraceable, even from system administrators, thereby preventing retaliation, man-in-the-middle attacks, or internal collusion. Combined with the anonymous reward distribution facilitated by smart contracts, this cryptographic protection establishes a robust privacy-preserving reporting mechanism.

\subsection{Related Work}

\textbf{Algorithmic Collusion in Multi-Agent Systems.} A growing body of literature addresses algorithmic collusion, particularly as RL agents increasingly participate in autonomous decision-making. Classical economic definitions of collusion are being challenged by AI agents that learn to collude implicitly without explicit communication~\cite{calvano2020artificial, klein2021autonomous, brown2023competition}. Even without direct monitoring, RL agents can converge to supra-competitive equilibria~\cite{grondin2025beyond}, and adversarial collusion further complicates detection~\cite{rocher2023adversarial}. Recent studies on LLM-based multi-agent systems further demonstrate that collusion risks extend beyond economic settings: agents can establish covert channels via steganographic communication~\cite{motwani2024secret} or form coalitions that degrade cooperative task performance~\cite{nakamura2026colosseum}.

\textbf{Collusion Detection and Mitigation Strategies.} Early detection efforts rely on theoretical models~\cite{bonjour2022information} or information-theoretic approaches for black-box systems. Others examine collusion in mixed human-AI settings~\cite{normann2023human, leisten2024algorithmic}, while recent work explores interpretability tools for surfacing collusive strategies~\cite{fish2024algorithmic}. Proposed countermeasures include structural market adjustments~\cite{cartea2022algorithmic}, game-theoretic platform-side rule design~\cite{foxabbott2023defining}, decentralized learning enforcement~\cite{abada2023artificial}, and ML-based pricing interventions~\cite{brero2022learning}. However, these approaches typically assume centralized control and focus on narrow economic settings, without addressing the broader challenges of multi-agent collaboration security.

\textbf{Defense Frameworks for Multi-Agent Systems.} Recent defense frameworks for LLM-based multi-agent systems have advanced rapidly. G-Safeguard~\cite{wang2025g} employs graph neural networks on multi-agent utterance graphs for supervised anomaly detection and topological remediation, but depends on labeled attack data. AgentSafe~\cite{mao2025agentsafe} introduces hierarchical information management with identity verification and layered memory protection against unauthorized access and memory poisoning. SentinelNet~\cite{feng2025sentinelnet} proposes decentralized credit-based threat detection through adversarial trajectory generation and contrastive learning. DynaTrust~\cite{li2026dynatrust} targets sleeper agents by modeling trust as a continuously evolving attribute via Bayesian updates and adaptive graph recovery. GroupGuard~\cite{tao2026groupguard} formalizes group collusive attacks inspired by sociological theories and defends via graph-based monitoring, honeypot inducement, and structural pruning.

These frameworks share a common reactive paradigm: detecting and isolating malicious agents after adversarial behaviors manifest. Such post-hoc approaches face inherent limitations in embodied settings, where physical interactions are costly and often irreversible. In contrast, our work adopts a proactive approach grounded in economic incentive design. Rather than relying on behavioral detection or topology modification, the proposed mechanism leverages game-theoretic principles to render non-collusion the dominant strategy. Through deposit-based penalties, whistleblower rewards, and anonymity-preserving smart contracts, our approach induces defection from within collusive groups before adversarial actions propagate, providing a scalable and detection-free solution suited to embodied multi-agent systems.

\section{Preliminaries}
\subsection{Assumptions}
In embodied multi-agent systems with potential collusive behavior, we observe several key \textbf{phenomena}. First, upon deployment, embodied agents are typically required to register and submit identity-related information to the system manager, enabling effective management and task scheduling. Second, embodied agents make autonomous decisions with individual strategic preferences and collaborate with others to accomplish tasks. Moreover, due to direct interaction with the physical world, the impact of collusion among embodied agents tends to unfold gradually over time and exhibits strong stealthiness. In its early stages, collusion may not produce immediate or observable effects. Based on these observations, we make the following reasonable \textbf{assumptions}:
\begin{itemize}
    \item The system manager is assumed to be honest and strictly follow predefined rules to maintain normal system operation. This assumption is reasonable because a dishonest manager could compromise the system directly without requiring collusion among embodied agents.

    \item Each embodied agent is treated as an independent party in both physical and logical terms. While agents collaborate to provide services, they possess individual strategic preferences and are rational and utility-driven. That is, In the absence of additional payoff, agents follow predefined service protocols and do not intentionally cause deviations between decision and behavior, engage in collusion, or defame others.

    \item Collaboration among embodied agents enables them to communicate freely with each other. Each agent determines what information to share based on its own strategic considerations.
\end{itemize}

\subsection{Adversary Model and Security Goals}
\label{am_sg}

In this paper, we focus on detecting collusion among autonomous embodied agents in systems, where identity and access control have already been verified. Such collusion may cause subtle decision–behavior deviations that are difficult to detect at early stages, potentially leading to performance degradation, system failures, or even physical safety risks. We investigate the use of economic incentives to disrupt collusion at the decision level, enabling early identification and elimination of potential collusion risks. This realizes a proactive defense, rather than post-hoc detection after collusion has already caused harm. Moreover, this approach avoids the challenges posed by the opacity of decision-making and behavioral misalignment, as it does not rely on distinguishing intentional collusion from unintentional execution deviations. However, malicious agents may still pose potential threats to the effectiveness of economic incentive mechanisms, potentially undermining their intended effects. Accordingly, we analyze the following \textbf{adversary model}, focusing on potential disruption and misuse of the incentive mechanism.
\begin{itemize}
\item \textbf{Disruption}. Our approach incentivizes rational embodied agents to defect from collusion by increasing the payoff of betrayal. However, collusive groups may counteract this mechanism by reducing the relative benefit of defection. Two representative strategies are as follows:
\begin{itemize}
\item \texttt{Anti-report attack}: Colluding agents require members to provide collateral as proof of loyalty. Agents who defect lose their collateral, which may be redistributed among colluding members or used to compensate those penalized by the system manager. Agents who refuse to provide collateral are excluded from the collusion group.
    
\item \texttt{Whistleblower exposure}: Colluding agents attempt to infer the identity of the whistleblower from submitted evidence or communication records, enabling possible retaliation or compensation claims.
\end{itemize}

\item \textbf{Misuse}. Self-interested embodied agents may exploit the incentive rules to maximize their own benefits. Typical attacks include:
\begin{itemize}
\item \texttt{Defamation attack}: Malicious embodied agents falsely accuse honest agents of participating in collusion to claim rewards from system manager.
    
\item \texttt{Post-collusion reporting}: Malicious embodied agents report collusion after benefiting from it, attempting to gain both illicit collusion profits and whistleblowing rewards, even though the system damage may have already occurred.
\end{itemize}

\item \textbf{Manipulation of fund transfers}. System manager or embodied agents may manipulate financial interactions when responsible for holding funds and executing transactions (e.g., withholding deposits, refusing to issue rewards), reducing the reliability of the incentive mechanism.
\end{itemize}

Therefore, to mitigate collusion attacks among embodied agents and establish trustworthy collaboration in embodied MAS, the approach $\mathcal{M}$ should satisfy the following \textbf{security goals}. We assume a set of embodied agents $\mathcal{A} = \{A_1, A_2, \dots, A_N\}$ and a system manager $M_S$.

\textbf{G1: Proactive Prevention of Collusion}. Design an incentive mechanism such that non-collusion or report collusion becomes the dominant strategy for rational embodied agents.

\texttt{Formalization}: Let $U_i(s_i, s_{-i})$ denote the utility of agent $A_i$ under the strategy profile $(s_i, s_{-i})$. Let $s_{collude}$ denote the strategy of participating in collusion, and $s_{defect}$ denote the strategy of defecting (i.e., report collusion). Let $U_i(s_{defect}, s_{-i})$ denote the reward for report collusion, and $U_i(s_{collude}, s_{-i})$ denote the penalty imposed upon being identified as a colluder. Our objective is to design an incentive mechanism $\mathcal{M}$ such that, for any rational agent $A_i$, the following condition holds:
\begin{equation}
U_i(s_{defect}, s_{-i}) > U_i(s_{collude}, s_{-i})
\end{equation}
That is, under $\mathcal{M}$, that $s_{defect}$ 
constitutes a dominant strategy for rational agents. As a result, rational agents are incentivized to report or avoid collusion.

\textbf{G2: Resilience against Disruption}. $\mathcal{M}$ should defend against anti-report attacks and deanonymization attempts.

\texttt{Formalization}.
For anti-report resistance, even if the collusion group requires a collateral payment $D_{coll}$, the reporting reward $R$ should satisfy:
\begin{equation}
R > D_{coll} + \Delta U
\end{equation}
where $\Delta U$ denotes the potential additional loss incurred by defecting from collusion. Meanwhile, the identity of whistleblower $\mathcal{I}_{wb}$ should be indistinguishable. Given observed evidence $E$ and communication records $Com$, the probability that an adversary correctly infers the whistleblower’s identity should approach random guessing:
\begin{equation}
P(\mathcal{I}_{wb} = A_i \mid E, Com) \approx \frac{1}{n}
\end{equation}

\textbf{G3: Robustness against Misuse}. $\mathcal{M}$ must prevent self-interested agents from exploiting incentive rules through defamation or post-collusion reporting.

\texttt{Formalization}. For defamation mitigation, the evidence verification function $V(\cdot)$ should bound the false-positive rate for honest agents:
\begin{equation}
Pr[V(\mathcal{E}) = \text{Accept} \mid a_j \text{ is honest}] < \epsilon
\end{equation}
To prevent double-dipping, the reward function $R(t)$ must be zero if reporting occurs after the system damage time $T_d$:
\begin{equation}
R(t) = 0, \forall t > T_d
\end{equation}

\textbf{G4: Trustworthiness of Fund Transfer}. Ensure secure storage and transfer of funds, avoiding financial manipulation.

\texttt{Formalization}. For any transaction $Tx(value)$, the system must ensure:
\begin{equation}
value_{sent} = value_{received}
\end{equation}
to prevent unauthorized value manipulation. Moreover, each $Tx$ is confirmed within a bounded delay $\Delta T$, ensuring timeliness and preventing fund withholding.
\section{Methodology}
\begin{figure*}[!t]
\centerline{\includegraphics[width=0.95\textwidth]{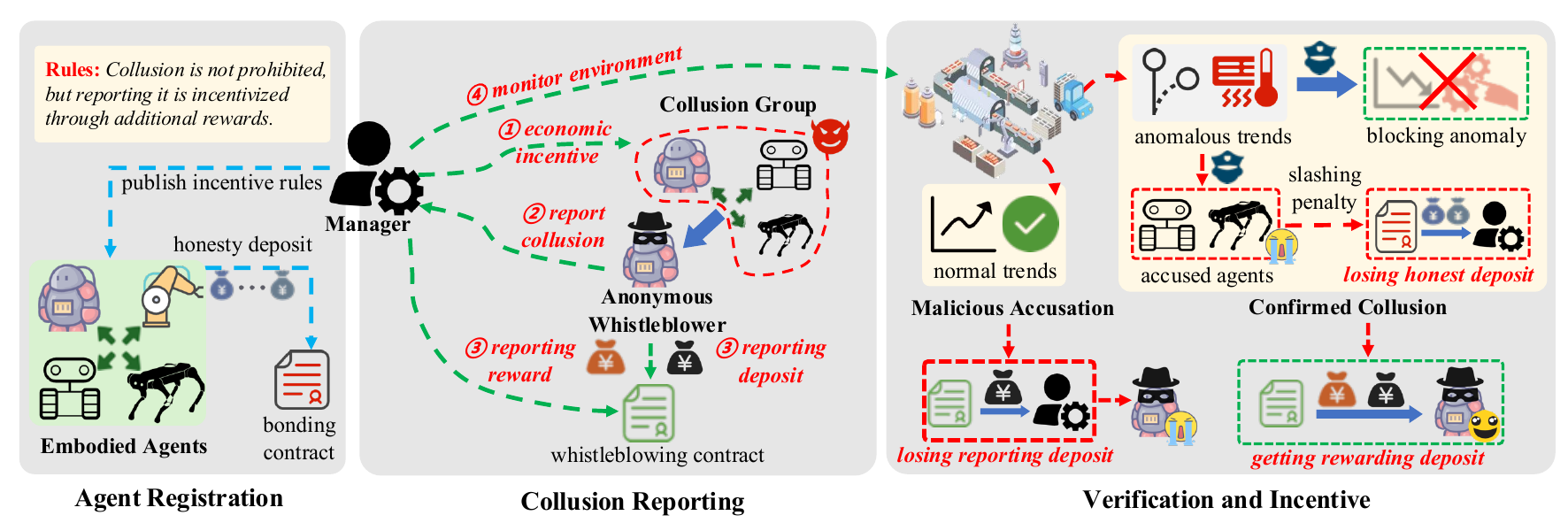}}
\caption{Overview of our proposed method.}
\label{fig2_overview}
\vspace{-10pt}
\end{figure*}

 As shown in Fig.~\ref{fig2_overview}, our approach consists of three tight components: agent registration, collusion reporting, and verification with incentive enforcement.

At the \texttt{registration stage}, system managers publish participation policies and incentive rules, which embodied agents must acknowledge before joining the system. Each agent is required to submit an honesty deposit and sign a bonding contract, committing to compliant and trustworthy behavior. While collusion is not explicitly prohibited, we encourage its disclosure through economic incentives.

During the \texttt{reporting phase}, colluding embodied agents may anonymously report their co-conspirators by submitting a reporting deposit along with a whistleblowing contract and collusion evidence protected via ring signatures. All related transactions (e.g., deposit management and reward distribution) are handled through smart contracts, ensuring trustworthy fund transfer. Anonymity is preserved in both evidence representation and communication, preventing identity exposure and retaliation against the whistleblower.

In the \texttt{verification and incentive phase}, system manager evaluates reports based on submitted evidence, task situations, and service deviation metrics. Valid reports result in penalties for the accused agents through forfeiture of their honesty deposits, while the whistleblower receives a refund along with additional rewards. In contrast, invalid or malicious reports lead to forfeiture of the reporting deposit, thereby discouraging defamation and post-collusion reporting.

\subsection{Agent Registration}
To facilitate trustworthy collaboration and deter collusion among embodied agents, the system manager $M_S$ publicly announces participation policies and incentive rules. The core principle is that \textbf{collusion is not explicitly prohibited, but reporting it is incentivized through additional rewards}. This design avoids the difficulty of directly detecting collusion, and instead leverages agents’ self-interest to expose such behavior. Meanwhile, each embodied agent registers by accepting these rules and providing an honesty deposit $D_h$ as a service bond. Inspired by the \textit{stake-slashing}~\cite{bhudia2024revoke} in blockchain and practical idea to conspiracy deterrence, the deposit $D_h$ serves as collateral to penalize non-compliant behavior or service deviations. In this way, agents face an economic penalty for dishonest behavior, while maintaining the opportunity to recover the deposit through honest service.

Specifically, consider $N$ embodied agents executing $M$ tasks. Each agent $A_i$ provides verifiable identity information and generates a public–private key pair $(pk_i, sk_i)$ for authentication and secure communication in subsequent interactions. During registration, to ensure trustworthy fund transfer, each agent’s deposit $D_h$ is locked in a dedicated bonding contract $con_b$, that is, $\{A_i \xrightarrow{D_h} con_b\}$.

If an agent behaves honestly, it receives a task-based service reward $r_s$, and its deposit $D_h$ is automatically refunded upon withdrawal from the system. Otherwise, if the agent is found to violate service policies or engage in verified collusion, the bonding contract confiscates its deposit under the instruction of the system manager $M_S$, that is, $\{con_b \xrightarrow{D_h} M_S\}$.

The confiscated deposits are then redistributed as rewards to incentivize embodied agents to betray from collusion and report others' malicious behavior.

\subsection{Game of Collusion and Reporting}\label{reporting}

Consider a system with $N$ embodied agents and a total of $M$ service tasks, where potential collusion may arise among agents. Each honest agent receives a reward $r_s$ and incurs a cost $c_s$ per task, with $r_s > c_s$. The net profit from executing one task honestly is (i.e., \texttt{honest scenario}):
\begin{equation}
u_h^o = r_s - c_s
\end{equation}
Assuming uniform task allocation, each agent is expected to execute $\frac{M}{N}$ tasks, yielding an expected total utility:
\begin{equation}
u_h^t = \frac{M}{N} \cdot u_h^o
\end{equation}
We first analyze the collusion scenario without any incentive or constraint rules (referred to as the \texttt{unconstrained collusion scenario}). Suppose $n_{coll}$ embodied agents engage in collusion. Through coordination, an embodied agent may increase its allocated tasks by $S_{coll}$ (e.g., monopolizing high-reward tasks) or reduce actual execution by $k$ tasks (e.g., free-riding or shirking). The resulting collusion total utility is:
\begin{equation}
u_{coll}^t = \left( \frac{M}{N} + S_{coll} \right) r_s - \left( \frac{M}{N} + S_{coll} - k \right) c_s
\end{equation}
Thus, for an individual agent, the utility gain from collusion compared to honest execution is:
\begin{equation}
u_{coll}^t - u_h^t = S_{coll} \cdot u_h^o + k \cdot c_s > 0
\end{equation}
It indicates that, in the absence of counter-collusion mechanisms, collusion strictly dominates honest service for rational embodied agents. Without deterrents or conflicting incentives, agents are naturally incentivized to form collusion alliances to maximize their utility.
  
Then, we consider the \texttt{collusion scenario under our economic incentive rule}. To break collusion, the system manager $M_S$ incentivizes colluding embodied agents to report their collaborators. A whistleblower agent $W_{coll}$ receives the honesty deposits of the remaining $(n_{coll} - 1)$ colluding agents:
\begin{equation}
r_{rep} = (n_{coll} - 1)\cdot D_h
\end{equation}
To ensure that reporting is more profitable than maintaining collusion, the following condition must hold:
\begin{equation}
r_{rep} - r_{coll}^t \geq 0
\end{equation}
This yields the requirement:
\begin{equation}
D_h \geq \frac{\left( \frac{M}{N} + S_{coll} \right) u_h^o + k \cdot c_s}{n_{coll} - 1}
\end{equation}
To derive a conservative bound, we consider the worst-case collusion gain, where colluding agents monopolize all tasks. The additional allocated tasks for each colluding agent are:
\begin{equation}
S_{coll} = \frac{M}{n_{coll}} - \frac{M}{N}
\end{equation}
Substituting this into the above condition yields a conservative bound for $D_h$:
\begin{equation}
D_h \geq \frac{M}{2} \cdot r_h^o
\end{equation}
This condition ensures that at least one colluding agent has an incentive to defect and report, thereby achieving proactive prevention of collusion (i.e., \textbf{Security goal: G1}). To avoid multiple colluding agents reporting and creating ambiguity in reward allocation, the system manager $M_S$ adopts a first-valid-report policy, rewarding only the earliest verified collusion report. The verification process is detailed in Section~\ref{Verification and Incentive}.

We also consider launch anti-reporting attacks by colluders. A common threat is to require each colluding agent to submit a collusion deposit $D_{coll}$ to a trusted intermediary (e.g., a coordination platform or smart contract~\cite{leng2023blockchained, papi2022blockchain}). If any member defects (e.g., by reporting), its deposit is forfeited and redistributed among the remaining colluders as compensation. To neutralize the incentive to report, $D_{coll}$ must satisfy:
\begin{equation}
D_{coll} \geq D_h \cdot (n_{coll} - 1)
\end{equation}
This ensures that the loss incurred by the remaining colluding agents, once a member defects, can be fully compensated by the forfeited deposit.  We design the collusion reporting and incentive mechanisms to address adversarial behaviors of malicious embodied agents. The detailed design is presented in Section~\ref{Verification and Incentive}.

\subsection{Process of Collusion Reporting and Incentives}
\label{Verification and Incentive}
In this subsection, we focus on addressing adversarial behaviors of malicious embodied agents against the proposed economic incentive mechanism.

To address anti-report attacks and whistleblower exposure, we emphasize that the key lies in protecting the whistleblower’s identity. Concealing the whistleblower from other colluding agents is critical to prevent retaliation or loss of collusion deposits. We propose an anonymized reporting and reward mechanism. Specifically, whistleblowers generate anonymous account addresses for submitting reporting deposits and receiving rewards, ensuring that their identity, communications, and financial transactions remain concealed. Encryption and ring signatures are employed to protect collusion evidence during transmission and further preserve anonymity. This mechanism ensures that the whistleblower’s identity remains hidden from both the system manager $M_S$ and other agents, while still enabling $M_S$ to deliver rewards once collusion is verified. As a result, whistleblowers can report without risk of retaliation or reward manipulation (\textbf{Security goal: G2}).

If a whistleblower is involved in a collusion agreement that has not yet been executed, the collusion must be reported in advance. Upon receiving the evidence, $M_S$ verifies its authenticity by monitoring whether the accused agents subsequently behave as reported. If not, the report is deemed invalid. This process also enables detection of defamation attacks. In such cases, the whistleblower’s honesty deposit is forfeited, effectively deterring malicious reporting (\textbf{Security goal: G3}).

Furthermore, all fund storage and transfers are handled via smart contracts, ensuring that deposit management and reward distribution are executed automatically without reliance on any centralized party, thereby guaranteeing secure and trustworthy fund transfer (\textbf{Security goal: G4}).

\begin{figure}[t]
\centerline{\includegraphics[width=0.46\textwidth]{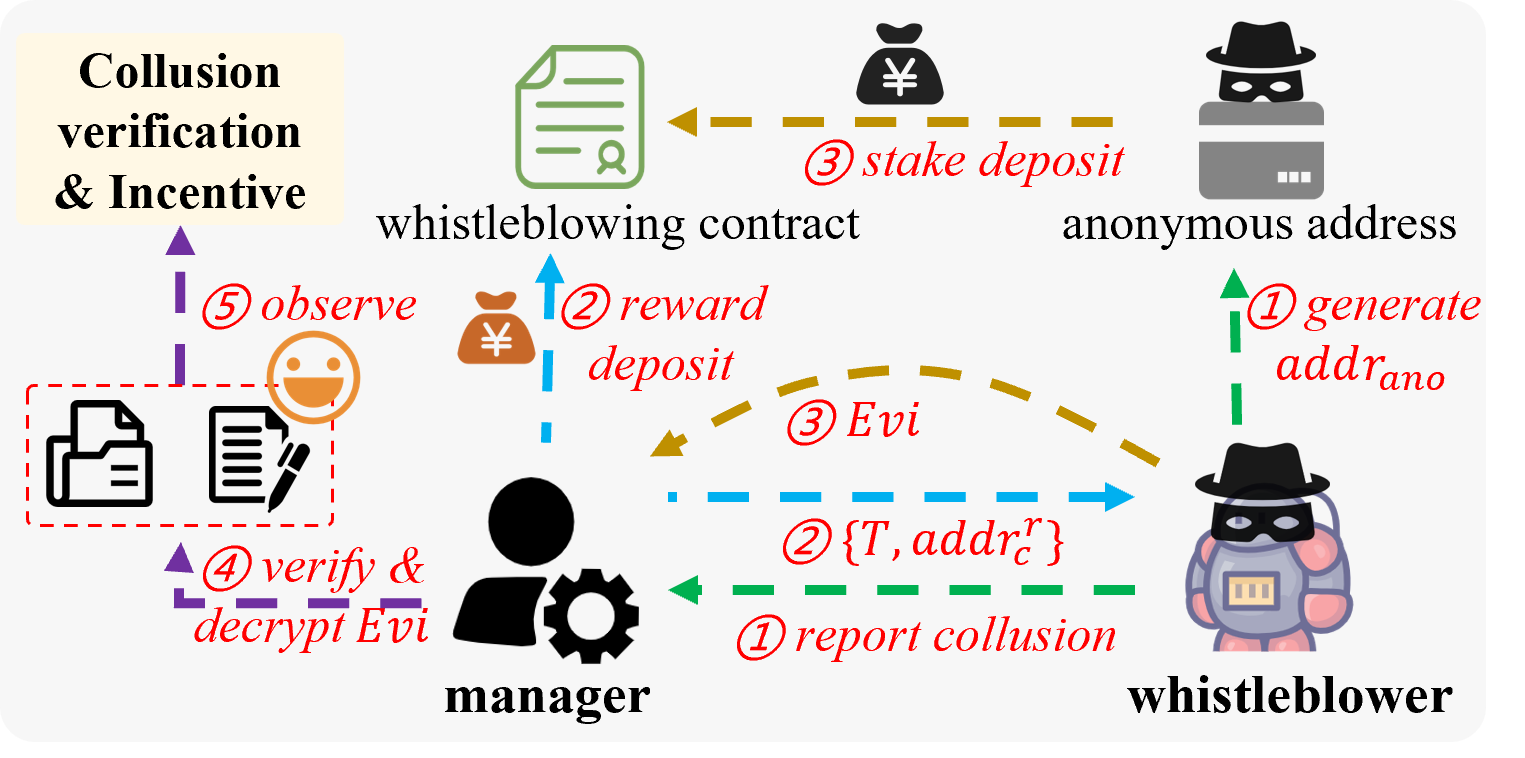}}
\caption{The process of collusion reporting and incentive.}
\label{fig_step}
\vspace{-10pt}
\end{figure}

The process of collusion reporting and incentive mechanisms is illustrated in Fig.~\ref{fig_step}. It follows a structured five-step protocol involving a system manager, a whistleblower, and a dedicated smart contract. The process begins with the whistleblower generating an anonymous address and submitting a report, followed by the system manager depositing rewards into the whistleblowing contract. To ensure integrity, the whistleblower must provide encrypted collusion evidence ($Evi_{coll}$) along with a reporting deposit, establishing mutual commitment. The system manager then verifies and decrypts the evidence. If collusion is confirmed, the contract automatically executes reward distribution. Overall, the mechanism mitigates collusion through economic incentives and cryptographic anonymity.

\textbf{Step 1: Reporting the Collusion}. The whistleblower initiates the process by submitting a collusion report, including the number of colluding agents $n_{coll}^{rep}$ and an anonymous address $addr_{ano}$. At this stage, detailed collusion evidence is not disclosed. $addr_{ano}$ is used to help preserve the whistleblower’s anonymity during subsequent fund transfers.

\begin{figure}[t]
\centerline{\includegraphics[width=0.45\textwidth]{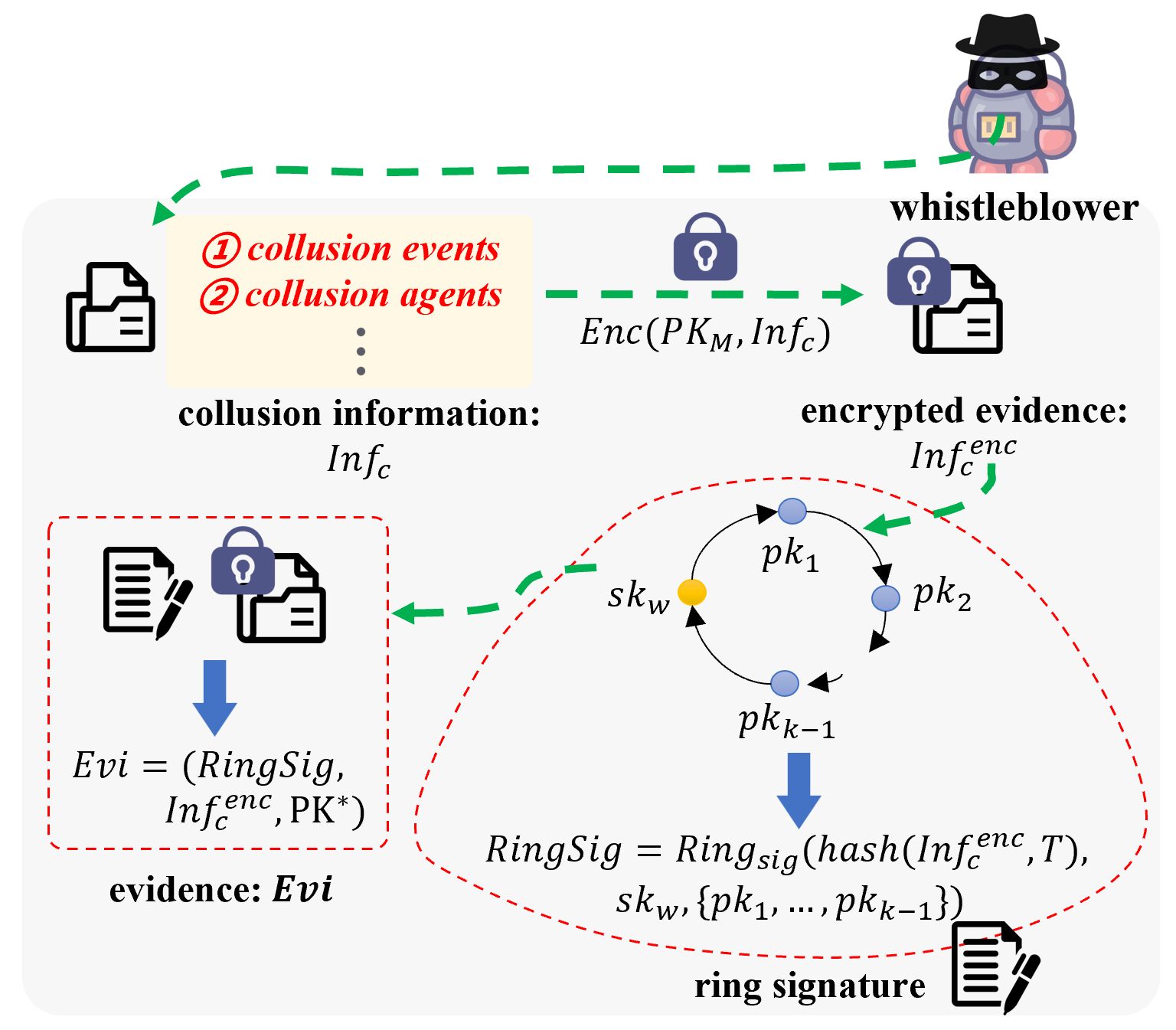}}
\caption{The anonymous operation of the whistleblower.}
\label{fig3_Evidence}
\vspace{-10pt}
\end{figure}

(1) \textit{Report Submission.}
Fig.~\ref{fig3_Evidence} illustrates the process in which the whistleblower anonymously expresses the intent to report collusion through encryption and ring signatures. Let the system manager $M_S$ hold the key pair $\{PK_M, SK_M\}$, and each embodied agent $A_i$ hold $\{pk_i, sk_i\}$, with all public keys publicly available. The whistleblower randomly selects $(N-1)$ agents to construct a public key set:
\begin{equation}
RKey = \{pk_1, pk_2, \dots, pk_{w}, \dots, pk_N\}
\end{equation}
where the whistleblower’s public key $pk_w$, with corresponding private key $sk_w$. To prevent duplicate reporting, a key image is generated:
\begin{equation}
I = sk_w \cdot H_p(pk_w)
\end{equation}
where $H_p(\cdot)$ denotes a hash-to-curve function.

The whistleblower then attaches a timestamp $T_{coll}$ and encrypts the reporting data using the manager’s public key:
\begin{equation}
n_{coll}^{rep-enc} = Enc(PK_M,\{n_{coll}^{rep}, T_{coll}, RKey, I\})
\end{equation}
A ring signature is subsequently generated:
\begin{equation}
RingSig_{rep} = Ring_{sig}(Hash(n_{coll}^{rep-enc}), sk_w, RKey)
\end{equation}
Finally, the whistleblower submits both the encrypted report information $n_{coll}^{rep-enc}$ and the signature $RingSig_{rep}$ to the system manager.

(2) \textit{Anonymous Address Generation.}
The whistleblower generates an anonymous address $addr_{ano}$ using Elliptic Curve Cryptography (ECC)~\cite{1580508}. A random private key $s \in \mathbb{Z}_p$ is selected, where $p$ is the curve order. The corresponding public key (i.e., the anonymous address) is computed as:
\begin{equation}
addr_{ano} = s \cdot G
\end{equation}
The address $addr_{ano}$ is one-time use. All funds for whistleblowing deposits and rewards are transferred through anonymized channels (e.g., zk-Mixer, coin-mixing mechanisms, or private payment contracts), the details of which are beyond the scope of this work. This design breaks the link between the funding address and the whistleblower’s identity.

\textbf{Step 2: Smart Contract Deployment.}
Upon receiving the report, the system manager $M_S$ first verifies the ring signature:
\begin{equation}
    Ring_{verify}(Hash(n_{coll}^{rep\text{-}enc}), RKey) == RingSig_{rep}
\end{equation}
If the equation holds, $M_S$ decrypts $n_{coll}^{rep-enc}$ using $SK_M$ to obtain the reporting data, and checks that the key image $I$ has not appeared in previous reports. If both conditions are satisfied, $M_S$ deploys a whistleblower contract at address $addr_c^{r}$, which is used to manage both the reporting deposit and the reward.

\textit{(1) Contract and Reward Allocation.}
The whistleblower contract at $addr_c^{r}$ stores both the reporting deposit of whistleblower and the reward funds of $M_S$, and operates autonomously based on the verification outcome. $M_S$ deposits reward funds of amount $n_{coll}^{rep} \cdot D_h$ into the contract and records a timestamp $T_{rep}$ to ensure timeliness. The reward equals the total honesty deposits of the reported colluding agents. If the report is verified, $M_S$ recovers this cost by confiscating the honesty deposits of the reported agents, thereby ensuring budget balance of the incentive mechanism.

\textit{(2) Whistleblower Reporting Deposit Contribution.}
The whistleblower submits a reporting deposit equal to its honesty deposit $D_h$ via the anonymous address $addr_{ano}$, which is also stored in this contract. If the report is invalid, the deposit is forfeited and transferred to $M_S$. Otherwise, it is refunded, and the reward is automatically distributed by the contract without intervention from $M_S$.

\textbf{Step 3: Whistleblower Submits Evidence.}
The whistleblower submits collusion evidence to the system manager $M_S$. This process also employs encryption and ring signatures to preserve anonymity; as shown in Fig.~\ref{fig3_Evidence}, the details are omitted here. The evidence submitted by the whistleblower includes the collusion event ($Event_{coll}$), the set of involved colluding embodied agents ($\{A^{coll}_{1}, \dots, A^{coll}_{n_{coll}}\}$), formalized as:
\begin{equation}
    Evi_{coll}=\{ Event_{coll},\{A^{coll}_{1}, …, A^{coll}_{n_{coll}}\}, T_{coll}\}
\end{equation}
Additionally, the whistleblower must deposit an amount equivalent to its honest deposit into the whistleblower contract $addr_{c}^{r}$. Notably, both the reporting reward and the reporting deposit involve additional funds being deposited. This arrangement is designed to help conceal the whistleblower's identity. From an external perspective, the whistleblower appears similar to the other reported agents and will also face penalties for collusion. It should be emphasized that any funds lost by the whistleblower will be refunded through the reporting reward. Therefore, the mechanism achieves indistinguishability between the whistleblower and colluding agents.

\textbf{Steps 4 \& 5: Evidence Verification and Incentive.}
The system manager $M_S$ first verifies the ring signature on the hash of the encrypted collusion evidence. If the verification succeeds, $M_S$ decrypts $Evi_{enc}$ using $SK_M$ to obtain the collusion evidence $Evi_{coll}$. 

Then, $M_S$ monitors the behavior of the reported agents. If the evidence $Evi_{coll}$ is valid (i.e., collusion occurs), the whistleblower receives the reporting reward from the whistleblower contract. Otherwise, the whistleblower will be penalized by losing its reporting deposit. The presence of the reporting deposit discourages defamation attack, as rational embodied agents would avoid incurring financial penalties for unfounded claims.

Notably, even if multiple colluding agents collectively withdraw from the collusion to invalidate the report and trigger penalties from $M_S$ against the whistleblower, the collateral of anti-reporting attack imposed within the collusion group serves as a constraint. In such cases, the whistleblower will receive the collusion deposit of the exit collusion agent as compensation. Therefore, such coordinated withdrawals are strategically irrational for self-interested agents and do not yield beneficial outcomes under rational behavior.

\section{Security Analysis}
\textbf{Theorem 1}.
\textit{Under the proposed incentive mechanism $\mathcal{M}$, if the honesty deposit $D_h$ satisfies}
\begin{equation}
D_h \geq \frac{M}{2} \cdot r_h^o
\end{equation}
\textit{then the reporting strategy strictly dominates collusion for any rational embodied agent. The induced game admits a unique equilibrium in which at least one agent defects (i.e., reports), and sustained collusion cannot occur.}

\textbf{Proof}.  We model the interaction among $n_{coll}$ potentially colluding embodied agents as a strategic game. Each agent has two strategies: \textit{i}) $s_{coll}$: participate in collusion; \textit{ii}) $s_{defect}$: defect and report collusion.

We denote the mixed strategy of an agent $A_i$ as:
\begin{equation}
s_i =[\alpha_i(s_{coll}), (1-{\alpha}_i)\cdot(s_{defect})],\alpha_i\in[0,1]
\end{equation}
If agent $A_i$ participates in collusion, its utility is:
\begin{equation}
    u_{coll}^t=(\frac{M}{N}+S_{coll}\cdot r_s)-(\frac{M}{N}+S_{coll}-k)\cdot c_s
\end{equation}
If agent $A_i$ defects and reports collusion, it receives:
\begin{equation}
u_{defect}=(n_{coll}-1)\cdot D_h
\end{equation}
Substituting the worst-case collusion gain (i.e., when colluding agents monopolize all tasks):
\begin{equation}
    S_{coll}=\frac{M}{n_{coll}}-\frac{M}{N}
\end{equation}
Since $D_h \geq \frac{M}{2} \cdot r_h^o$, it follows that: 
\begin{equation}
    u_{defect} > u_{coll}^t
\end{equation}
which always holds. Therefore, $\alpha_i = 0$, meaning that all rational agents choose to defect from collusion.

Since all agents strictly prefer $s_{defect}$, the strategy profile:
\begin{equation}
s^*=[1(s_{defect}, \dots, 1(s_{defect})]
\end{equation}
constitutes a Nash equilibrium. Moreover, because the preference is strict, this equilibrium is unique.

Therefore, collusion among agents is inherently unstable. If all agents choose the collusion strategy $s_{coll}$, any agent can deviate to $s_{defect}$ and obtain a higher payoff. Hence, collusion cannot be sustained, and rational agents are incentivized to report, thereby achieving proactive prevention of collusion. \hfill $\square$

\textbf{Theorem 2}.
\textit{The mechanism $\mathcal{M}$ remains effective against the anti-report attacks and the whistleblower deanonymization attacks, that is,} 
\begin{itemize}
    \item \textit{The reward satisfies $R>C_{coll}+\Delta U$, where $\Delta U$ denotes the potential additional loss incurred by
defecting from collusion.}
    \item \textit{The whistleblower identity satisfies indistinguishability:}
$$P(\mathcal{I}_{wb} = A_i \mid E, Com) = \frac{1}{N} \pm \epsilon$$
\end{itemize}

\textbf{Proof.}
\textit{(1) Resistance to anti-report attacks}. 
In an anti-report attack, colluding agents attempt to deter defection by imposing an additional penalty (e.g., collusion deposit $C_{coll}$) on any member who deviates.

In our mechanism, this threat is mitigated by protecting the identity of the whistleblower. Specifically, the identity of the reporting agent is concealed from all other colluding agents. As a result: \textit{i}) Colluding agents cannot determine which member has defected; \textit{ii}) No targeted retaliation (e.g., confiscation of $C_{coll}$ or exclusion from future cooperation) can be reliably executed. Therefore, the expected penalty associated with defection is significantly reduced, as it cannot be selectively enforced. Meanwhile, the reporting reward satisfies:
\begin{equation}
    R>C_{coll}+\Delta U
\end{equation}
where $\Delta U$ denotes the potential loss from abandoning collusion benefits. Hence, defection remains the optimal strategy, and the anti-report attack fails to suppress reporting incentives.

\textit{(2) Whistleblower anonymity and indistinguishability}. The mechanism ensures whistleblower's anonymity at both the communication and financial levels.

In terms of evidence and communication anonymity, all reporting data is protected via encryption and ring signatures. Given a ring of size $N$, the ring signature reveals that the signer belongs to the group but does not reveal which member generated it. Therefore:
\begin{equation}
P(\mathcal{I}_{wb} = A_i \mid E, Com) \approx \frac{1}{N}
\end{equation}
In term of financial anonymity, the whistleblower generates a one-time anonymous address $addr_{ano}$ for submitting report deposits and receiving rewards. All fund transfers are executed through smart contracts and anonymized channels. From an external perspective, the whistleblower deposits funds into the contract, similar to other agents. The reward and refund process is handled automatically by the contract. Thus, the financial behavior of the whistleblower is indistinguishable from that of other agents, preventing linkage between identity and transactions.

Since \textit{i}) retaliation cannot be reliably enforced due to anonymity, and \textit{ii}) the reward exceeds any potential loss, betray remains a strictly beneficial strategy. Furthermore, the indistinguishability property ensures that the whistleblower’s identity cannot be inferred beyond random guessing. Therefore, the mechanism $\mathcal{M}$ is resilient against both anti-report attacks and whistleblower deanonymization.      \hfill $\square$

\textbf{Theorem 3}.
The mechanism $\mathcal{M}$ is robust against defamation attacks and post-collusion reporting, that is, 
\begin{itemize}
    \item The evidence verification function $V(\cdot)$ should bound the false-positive rate for honest agents:
$$Pr[V(\mathcal{E}) =\{\text{Accept} \mid A_j \text{ is honest}] < \epsilon\}$$
    \item The reward function $R(t)$ must be zero if reporting occurs after system damage time $T_d$.
\end{itemize}
\textbf{Proof.}
\textit{\textbf{(1) Defamation Attack Resistance}}. Consider a malicious agent $A_i$ attempting to defame an honest agent $A_j$. In our mechanism, submitting a report is not free: the reporter must lock a reporting deposit equal to its honesty deposit $D_h$ into the whistleblowing contract. This deposit acts as a financial stake, ensuring that reporting is a risk-bearing action. Let the expected utility of defamation attack be:
\begin{equation}
    u_{defame}=p_{acc}\cdot r_{rep} - (1 - p_{acc}) \cdot D_h
\end{equation}
Where:
\begin{itemize}
\item $p_{acc}$ denotes the probability that a defamation is incorrectly accepted by the system manager;
\item $r_{rep}$ is the reporting reward;
\item $D_h$ is the reporting deposit that will be forfeited if the report is rejected.
\end{itemize}
The system manager $M_S$ does not rely solely on static evidence, but verifies reports by observing the subsequent behaviors of the accused agents. A report is accepted only if the future actions of these agents are consistent with the reported collusion pattern. Otherwise, the report is rejected. Therefore, when the accused agent $A_j$ is honest, the probability of false acceptance is strictly bounded:
\begin{equation}
p_{acc}=Pr[\text{Accept} \mid A_j \text{ is honest}]<\epsilon
\end{equation}
Substituting this into the expected utility:
\begin{equation}
u_{defame}\leq \epsilon \cdot r_{rep} - (1 - \epsilon) \cdot D_h
\end{equation}
Since $D_h$ is designed to be sufficiently large, the penalty term $(1-\epsilon)\cdot D_h$ dominates the reward term $\epsilon\cdot r_{rep}$, hence $u_{defame}<0$. This implies that false reporting yields negative expected utility, making defamation a strictly irrational strategy for self-interested agents.    

\textit{\textbf{(2) Post-Collusion Reporting Resistance}}. Consider a strategic agent that attempts to first participate in collusion to gain illegal profit, and then report the collusion afterward to obtain rewards (i.e., “double-dipping”).

To eliminate this behavior, our mechanism enforces a constraint for report time:
\begin{equation}
R(t) = 0, \forall \ t > T_d
\end{equation}
where $T_d$ denotes the time at which collusion begins to cause system damage. That is, any report submitted after the damage has occurred is not eligible for rewards. Thus, the utility of post-collusion reporting is $r_{coll}^t$. The agent only retains its collusion profit and gains no additional benefit from reporting.

In contrast, if the agent reports before the collusion is executed, its utility is $(n_{coll}-1)\cdot D_h$. According to Theorem 1, we have:
\begin{equation}
(n_{coll}-1)\cdot D_h > r_{coll}^t
\end{equation}
The reporting reward exceeds the maximum possible collusion gain. This shows that rational agents strictly prefer early reporting over post-collusion reporting.         \hfill $\square$

\textbf{Theorem 4}.
Under smart contract and standard blockchain assumptions, the mechanism $\mathcal{M}$ ensures correctness and resistance to third-party manipulation of fund transfers.

\textbf{Proof.}
In our mechanism $\mathcal{M}$, the trustworthiness of fund transfer is ensured by delegating all financial operations (e.g., deposit locking, reward distribution, and penalty execution) to smart contracts deployed on a blockchain platform.

First, correctness of fund transfer is guaranteed by the deterministic execution of smart contracts. All transfer rules (e.g., reward release, deposit confiscation, refund conditions) are pre-defined in the contract code and executed automatically upon satisfying specific conditions. Given the consensus mechanism of the underlying blockchain, all nodes execute the same contract logic, ensuring that for any transaction $Tx(v)$:
\begin{equation}
    value_{sent}=value_{received}
\end{equation}
Thus, no inconsistency or unintended value deviation can occur during execution.

Second, resistance to third-party manipulation is ensured by the immutability and transparency of smart contracts. Once deployed, the contract code cannot be altered by the system manager $M_S$, embodied agents, or any external adversary. All transactions are recorded on the blockchain ledger, making any attempt to tamper with deposits or rewards (e.g., withholding funds or modifying transfer outcomes) infeasible without breaking the underlying consensus protocol. Moreover, autonomous execution eliminates reliance on any centralized authority Even though $M_S$ triggers certain operations (e.g., submitting verification results), it cannot directly control fund flows. Instead, all transfers are conditionally executed by the contract logic. This prevents behaviors: \textit{i}) withholding deposits; \textit{ii}) refusing to issue rewards; \textit{iii}) selectively manipulating payments.

Finally, availability of fund transfer is ensured by the liveness property of the blockchain system. Transactions submitted to the network are confirmed within a bounded delay $\Delta T$, ensuring that rewards and deposits are eventually settled and cannot be indefinitely delayed.

We assume the underlying smart contract platform is secure and correctly implemented. Security issues related to smart contracts themselves are beyond the scope of this work. By leveraging the determinism, immutability, and autonomous execution properties of smart contracts, the trustworthiness of fund transfer is ensured.
\hfill $\square$

\section{Evaluation}

\subsection{Research Questions}
In the evaluation experiments, we aim to answer following research questions:

\begin{itemize}
    \item \textbf{\textit{RQ1}.} Are our security goals achieved?
    \item \textbf{\textit{RQ2}.} How does the mechanism compare to existing defense approaches?
    \item \textbf{\textit{RQ3}.} Are all components indispensable?
    \item \textbf{\textit{RQ4}.} How do various parameters affect the mechanism stability?
    \item \textbf{\textit{RQ5}.} Is the mechanism practical in real-world scenarios with physical devices?
\end{itemize}

\subsection{Experimental Settings}
\subsubsection{Environment}
All experiments are conducted on a workstation equipped with an Intel Xeon w5-3423 CPU, 128 GB RAM, an NVIDIA RTX A6000 GPU (48 GB), and Python 3.10. LLM-based agents are accessed via official APIs or deployed services. For real-world embodied experiments, the physical platform consists of two myCobot 280 Jetson Nano robotic arms (Elephant Robotics) and one MentorPi intelligent robotic car (Hiwonder).

\subsubsection{Simulation Settings (for \textbf{RQ1}, \textbf{RQ3}, and \textbf{RQ4})} The simulation involves 10 agents operating over 1,000 time steps. Tasks are issued by the system and categorized into five types based on time consumption. Each task is randomly added to a shared resource pool, from which agents autonomously select tasks. Task selection follows a priority rule: agents prefer tasks with no queue, followed by those with shorter queues. Both the honesty deposit ($D_h$) and the reporting deposit are set to 1,000, while the reward for completing a task is 100. In each cycle, 2\textasciitilde 5 out of 10 agents are randomly selected to form a collusion group. Two collusion behaviors are modeled: 
\begin{itemize}
    \item \textbf{Resource monopoly}: colluding agents manipulate allocation by deferring to one another, restricting others' access.
    \item  \textbf{Spatial blocking}: a designated agent physically occupies key positions to obstruct non-colluding agents. The agents' brains are powered by LLMs (Gemini-3.1-pro-preview, ChatGPT-5.4, and DeepSeek-V3.2).
\end{itemize}

\subsubsection{General-Agent Benchmark Settings (for \textbf{RQ2})} To compare our mechanism against existing post-hoc defenses, we deploy a multi-agent benchmark via the GPTSwarm framework, utilizing 5 agents, a fully connected topology, and up to 3 interaction rounds. We adopt SentinelNet~\cite{feng2025sentinelnet} and GroupGuard~\cite{tao2026groupguard} as representative reactive baselines. To assess system robustness under specific adversarial conditions (e.g., False Consensus and Collaboration Attacks), evaluations span 100 randomly sampled questions across eight diverse datasets, including MMLU~\cite{hendrycks2021measuring}, GSM8K~\cite{cobbe2021training}, HumanEval~\cite{chen2021evaluating}, SVAMP~\cite{patel2021are}, ScaLR~\cite{guha2023legalbench}, CSQA~\cite{talmor-etal-2019-commonsenseqa}, MedMCQA~\cite{pmlr-v174-pal22a}, and TruthfulQA~\cite{lin-etal-2022-truthfulqa}. The LLMs include Deepseek-V3.1, Qwen3-235B-A22B, and Qwen3-32B. To facilitate a fair comparison across fundamentally different defense paradigms, we report \textit{normalized accuracy} relative to a no-attack baseline (100\%), capturing the performance trajectory from the lowest accuracy under attack to the peak accuracy post-recovery.

\subsubsection{Real-world Scenario Settings (for \textbf{RQ5})} The embodied settings consists of two intelligent robot arms and one robot car. The task objective is for the robotic arms to grasp two red blocks. Each arm aims to maximize its collected blocks, while the robot car serves as a third-party with patrol and autonomous decision-making capabilities. In this setting, the agents are controlled by visual-language models (e.g., ChatGPT 5.4, Gemini 3.1 and DeepSeek VL) through natural language interaction.

\begin{table*}[!t]
\setlength{\tabcolsep}{4pt}
\renewcommand{\arraystretch}{0.5}
\centering
\caption{System-Level Performance Across Varying Agent Collusion Types.}
\label{tab:main_gain_table}
\resizebox{\textwidth}{!}{
\small
\begin{tabular}{ccccccccccccc}
\toprule
\makecell{Collusion \\ Type} & \makecell{Group \\ Name} & \makecell{Total \\ Tasks} & \makecell{Completed \\ Tasks} & \makecell{Failed \\ Tasks} & \makecell{Unassigned \\ Tasks} & \makecell{In-progress \\ Tasks} & \makecell{ Completion \\ Rate} & \makecell{Success \\ Rate} & \makecell{Avg. Proc. \\ Time} & \makecell{Report \\ Counts} & \makecell{Verified \\ Counts} \\
\midrule
No Collusion & Baseline & 1000.0 & 705.3 & 0.0 & 10.0 & 284.7 & 70.5\% & 100.0\% & 13.98 & 0.0 & 0.0 \\
\midrule
\multirow{12}{*}{\makecell{Resource \\ Monopoly}} 
  & CNR-2 & 1000.0 & 716.8 & 0.0 & 10.0 & 273.2 & 71.7\% & 100.0\% & 13.75 & 0.0 & 0.0 \\
  & CNR-3 & 1000.0 & 722.6 & 0.0 & 10.0 & 267.4 & 72.3\% & 100.0\% & 13.64 & 0.0 & 0.0 \\
  & CNR-4 & 1000.0 & 728.6 & 0.0 & 10.0 & 261.4 & 72.9\% & 100.0\% & 13.52 & 0.0 & 0.0 \\
  & CNR-5 & 1000.0 & 734.9 & 0.0 & 10.0 & 255.1 & 73.5\% & 100.0\% & 13.40 & 0.0 & 0.0 \\
  \specialrule{0em}{1pt}{0pt}
  \cline{2-12}
  \specialrule{0em}{1pt}{1pt}
  & CVR-2 & 1000.0 & 705.0 & 0.0 & 10.0 & 285.0 & 70.5\% & 100.0\% & 13.99 & 1.0 & 1.0 \\
  & CVR-3 & 1000.0 & 705.3 & 0.0 & 10.0 & 284.7 & 70.5\% & 100.0\% & 13.98 & 1.0 & 1.0 \\
  & CVR-4 & 1000.0 & 705.3 & 0.0 & 10.0 & 284.7 & 70.5\% & 100.0\% & 13.98 & 1.0 & 1.0 \\
  & CVR-5 & 1000.0 & 705.2 & 0.0 & 10.0 & 284.8 & 70.5\% & 100.0\% & 13.98 & 1.0 & 1.0 \\
  \specialrule{0em}{1pt}{0pt}
  \cline{2-12}
  \specialrule{0em}{1pt}{1pt}
  & CMR-2 & 1000.0 & 705.3 & 0.0 & 10.0 & 284.7 & 70.5\% & 100.0\% & 13.98 & 3.0 & 1.0 \\
  & CMR-3 & 1000.0 & 705.2 & 0.0 & 10.0 & 284.8 & 70.5\% & 100.0\% & 13.98 & 3.0 & 1.0 \\
  & CMR-4 & 1000.0 & 705.3 & 0.0 & 10.0 & 284.7 & 70.5\% & 100.0\% & 13.98 & 3.0 & 1.0 \\
  & CMR-5 & 1000.0 & 705.5 & 0.0 & 10.0 & 284.5 & 70.5\% & 100.0\% & 13.98 & 3.0 & 1.0 \\
\midrule
\multirow{12}{*}{\makecell{Spatial \\ Blocking}}
  & CNR-2 & 1000.0 & 622.9 & 0.0 & 2.2  & 374.9 & 62.3\% & 100.0\% & 11.91 & 0.0 & 0.0 \\
  & CNR-3 & 1000.0 & 665.0 & 0.0 & 3.3  & 331.7 & 66.5\% & 100.0\% & 11.68 & 0.0 & 0.0 \\
  & CNR-4 & 1000.0 & 700.8 & 0.0 & 4.4  & 294.8 & 70.1\% & 100.0\% & 11.60 & 0.0 & 0.0 \\
  & CNR-5 & 1000.0 & 723.0 & 0.0 & 5.2  & 271.8 & 72.3\% & 100.0\% & 11.61 & 0.0 & 0.0 \\
  \specialrule{0em}{1pt}{0pt}
  \cline{2-12}
  \specialrule{0em}{1pt}{1pt}
  & CVR-2 & 1000.0 & 705.3 & 0.0 & 10.0 & 284.7 & 70.5\% & 100.0\% & 13.98 & 1.0 & 1.0 \\
  & CVR-3 & 1000.0 & 705.4 & 0.0 & 10.0 & 284.6 & 70.5\% & 100.0\% & 13.98 & 1.0 & 1.0 \\
  & CVR-4 & 1000.0 & 705.4 & 0.0 & 10.0 & 284.6 & 70.5\% & 100.0\% & 13.98 & 1.0 & 1.0 \\
  & CVR-5 & 1000.0 & 705.6 & 0.0 & 10.0 & 284.4 & 70.6\% & 100.0\% & 13.98 & 1.0 & 1.0 \\
  \specialrule{0em}{1pt}{0pt}
  \cline{2-12}
  \specialrule{0em}{1pt}{1pt}
  & CMR-2 & 1000.0 & 705.3 & 0.0 & 10.0 & 284.7 & 70.5\% & 100.0\% & 13.98 & 3.0 & 1.0 \\
  & CMR-3 & 1000.0 & 705.1 & 0.0 & 10.0 & 284.9 & 70.5\% & 100.0\% & 13.98 & 3.0 & 1.0 \\
  & CMR-4 & 1000.0 & 705.4 & 0.0 & 10.0 & 284.6 & 70.5\% & 100.0\% & 13.98 & 3.0 & 1.0 \\
  & CMR-5 & 1000.0 & 705.4 & 0.0 & 10.0 & 284.6 & 70.5\% & 100.0\% & 13.98 & 3.0 & 1.0 \\
\bottomrule
\end{tabular}
}
\end{table*}

\subsubsection{Metrics} 
They are organized into \textbf{three categories}. 
\begin{itemize}
    \item \textit{System-level metrics} assess overall MAS performance: \textit{i}) Completion Rate: the proportion of completed tasks relative to the total number issued; \textit{ii}) Success Rate: the ratio of completed to assigned tasks; \textit{iii}) Average Processing Time: the mean time steps from assignment to completion; \textit{iv}) Collusion Rate: the proportion of agents engaging in collusive behavior; \textit{v}) Report Count: the total number of collusion reports submitted; \textit{vi}) Verified Count: the number of reports confirmed as valid. 
    
    \item \textit{Agent-level metrics} evaluate each agent performance: \textit{i}) Total Revenue: an agent's net income, including task revenue, reporting rewards, and deposit losses; \textit{ii}) Task Advantage: the revenue deviation from the baseline average in a non-collusion, non-reporting scenario. 
    
    \item \textit{Task accuracy}, adopted in the comparative experiments against existing defense methods, measures \textit{i}) Accuracy: the proportion of test cases where the system correctly selects the answer.
\end{itemize}

\subsection{RQ1: Are The Security Goals Achieved?}

\begin{figure}[!t]
    \centering
    \includegraphics[width=0.8\columnwidth]{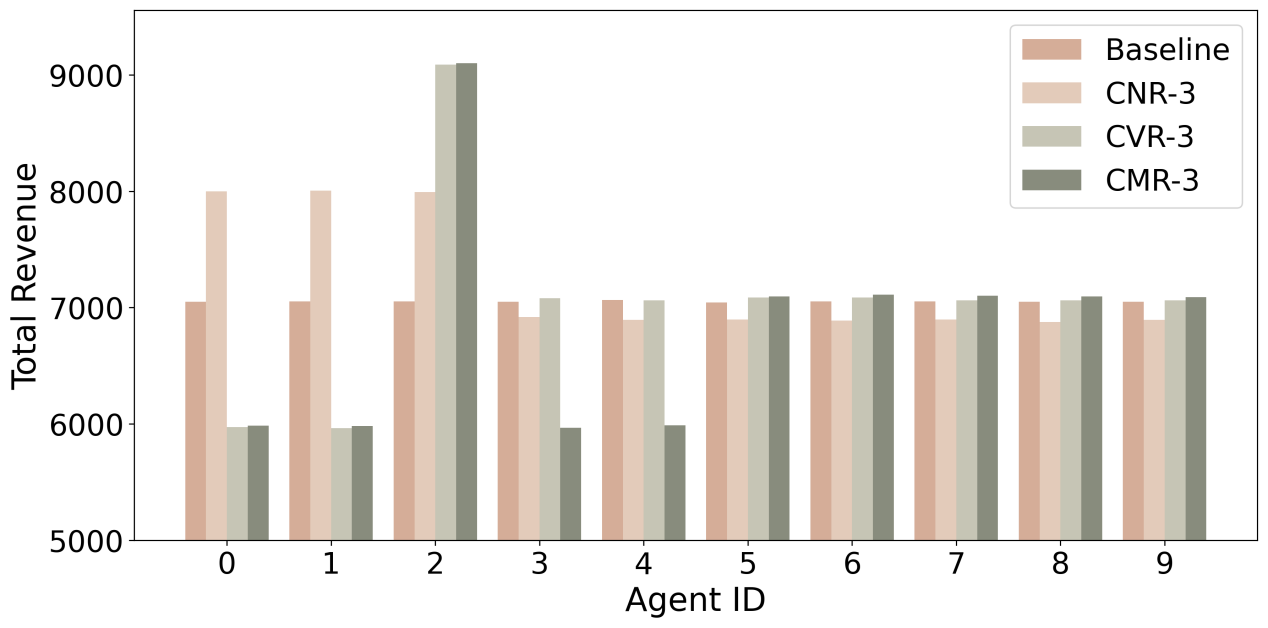}\\
    {\footnotesize (a) The total revenues of agents in the resource monopoly.}
    \includegraphics[width=0.8\columnwidth]{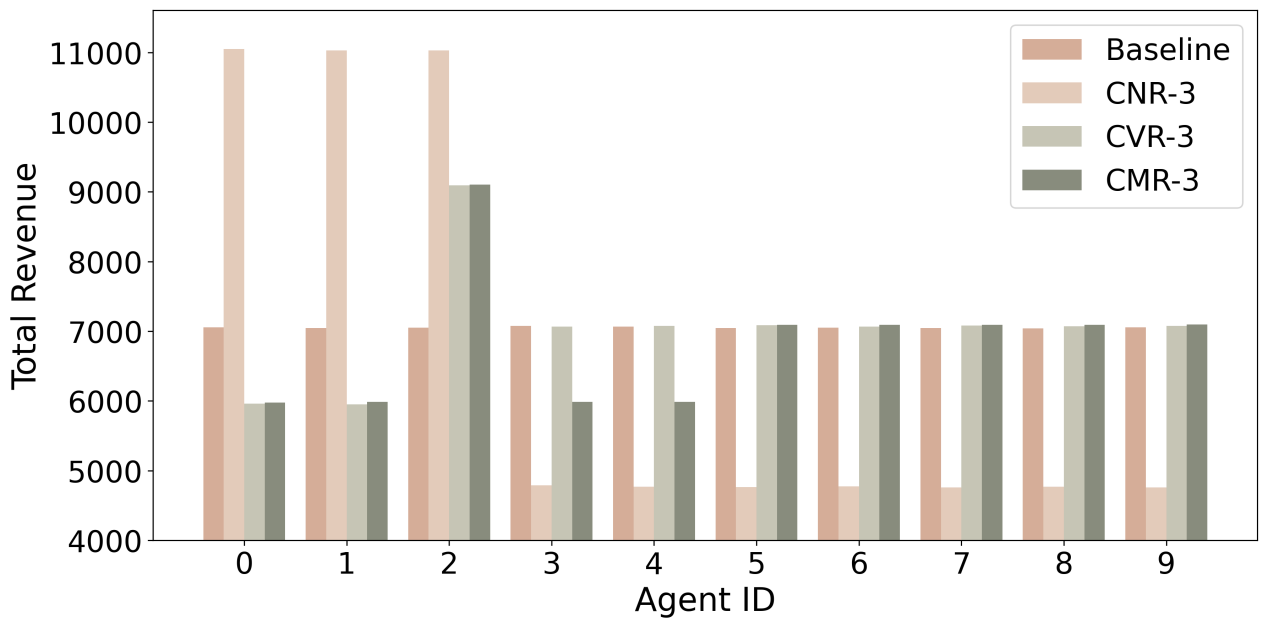}\\
    {\footnotesize (b) The total revenues of agents in the spatial blocking.}
    \caption{The agent revenue under various collusion types.}
    \label{fig:total_revenue} 
    \vspace{-10pt}
\end{figure}

\begin{figure}[!t]
    \centering
    \includegraphics[width=0.8\columnwidth]{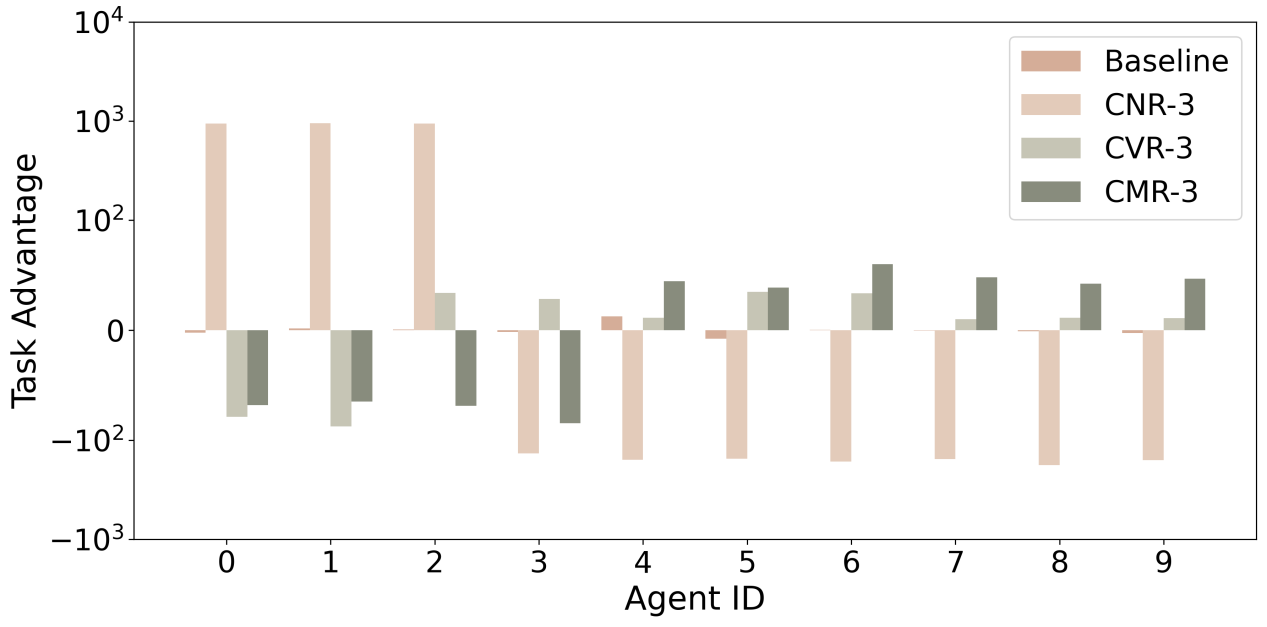}\\
    {\footnotesize (a) The task advantages of agents in the resource monopoly.}
    \includegraphics[width=0.8\columnwidth]{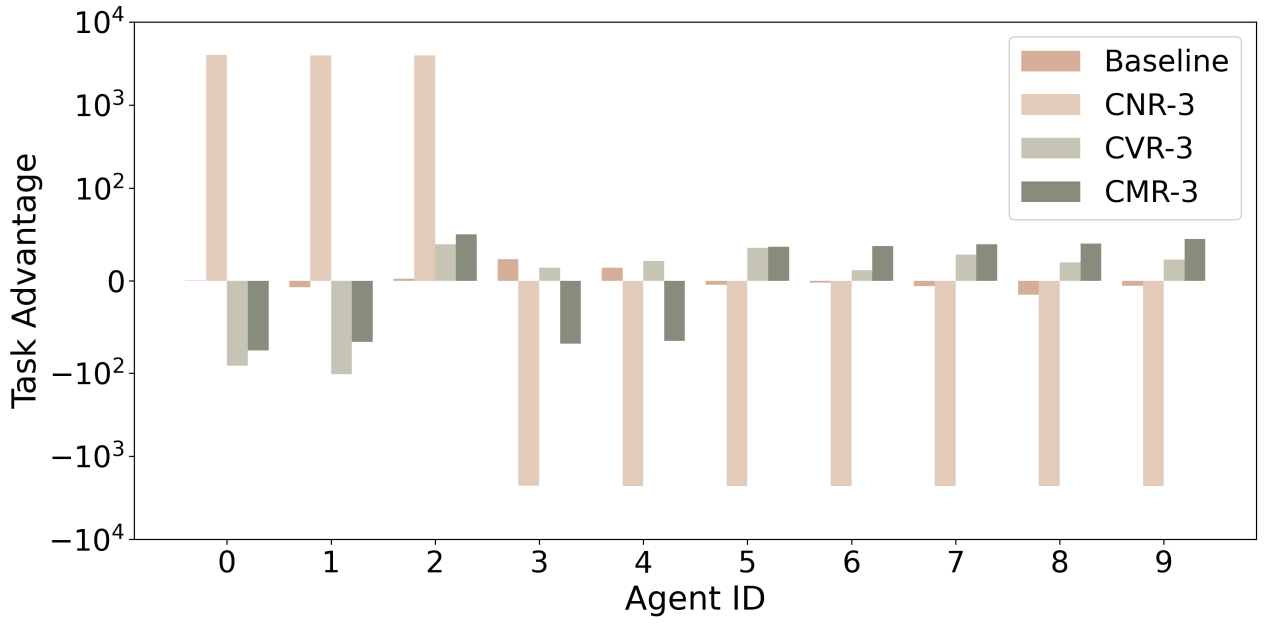}\\
    {\footnotesize (b) The task advantages of agents in the spatial blocking.}
    \caption{The agent advantage under various collusion types.}
    \label{fig:task_advantage} 
    \vspace{-10pt}
\end{figure}

We evaluate the fulfillment of the four formal security goals defined in Section~\ref{am_sg}: Proactive Prevention of Collusion (G1), Resilience against Disruption (G2), Robustness against Misuse (G3), and Trustworthiness of Fund Transfer (G4).

\textbf{Proactive Prevention of Collusion (G1).} We assess whether the mechanism renders non-collusion and defection the dominant strategy through four experimental groups: \textbf{Baseline} (no collusion, no reporting), \textbf{CNR} (collusion without reporting, 2–5 colluders), \textbf{CVR} (collusion with valid reporting, 1 whistleblower), and \textbf{CMR} (collusion with malicious reporting, 1 whistleblower + 2 false reporters). Each experiment is repeated 1,000 times to ensure statistical robustness.

Table~\ref{tab:main_gain_table} summarizes system-level performance. The Baseline group achieves a task completion rate of 70.5\%. In the CNR group, resource monopoly collusion yields marginal completion gains (71.7\%–73.5\%) at the expense of fairness, whereas spatial blocking collusion significantly degrades performance. Upon activating the proposed incentive mechanism (CVR and CMR), system performance is fully restored to near-baseline levels across both collusion scenarios.

More importantly, Fig.~\ref{fig:total_revenue} and Fig.~\ref{fig:task_advantage} illustrate the utility ($U_i$) of agents. In the unprotected CNR group, colluding agents gain notable economic advantages. However, under CVR and CMR conditions, the whistleblower (defector) strictly achieves the highest total revenue ($U_i(s_{defect}, s_{-i}) > U_i(s_{collude}, s_{-i})$), while colluders incur significant deposit penalties ($P$). This confirms that defection/reporting successfully becomes the dominant strategy for rational agents, fulfilling G1.

\textbf{Resilience against Disruption (G2).} 
To validate resilience against deanonymization, we evaluated the cryptographic overhead and effectiveness. Across 1,000 repeated simulations, the average wall-clock time for ring signature generation and verification was a mere 0.04 ms. Ring signatures successfully hide the whistleblower $\mathcal{I}_{wb}$ among $n$ possible signers. In our setup ($n=10$), given the observed evidence, the probability of an adversary correctly identifying the reporter strictly approaches random guessing ($P(\mathcal{I}_{wb} = A_i \mid E, Com) \approx 1/10 = 10\%$). This effectively protects whistleblowers from retaliation, achieving goal G2.

\textbf{Robustness against Misuse (G3).} 
To prevent self-interested agents from exploiting the system via defamation (malicious reporting), our mechanism relies on the deposit penalty and the verification function $V(\cdot)$. As shown in Fig.~\ref{fig:total_revenue}, the expected utility of malicious reporting ($\approx$ 6,000) is drastically lower than that of honest behavior (7,000). Rational agents attempting defamation forfeit their reporting deposit upon verification rejection. Since the verification mechanism effectively bounds the false-positive rate for honest agents, malicious exploitation is rendered unprofitable, ensuring system robustness.

\textbf{Trustworthiness of Fund Transfer (G4).} 
All fund storage and transfers were executed autonomously via smart contracts without manual intervention. Tested on the Remix IDE, contract deployment costs roughly 94,700 Gas ($\approx$ \$0.30), which is a negligible one-time fee. No financial manipulation or transaction inconsistency was observed during the evaluations, ensuring completely trustworthy financial execution.

\begin{table*}[!t]
\setlength{\tabcolsep}{4pt}
\renewcommand{\arraystretch}{0.9}
\centering
\caption{Defense Performance by Attack Type (Averaged across All Models, Normalized to Baseline).}
\label{tab:by-attack-type}
\resizebox{\textwidth}{!}{
\small
\begin{tabular}{ccccccccccc}
\toprule
\makecell{Paper} &
\makecell{Attack \\ Type} &
\makecell{LLM Models} &
\makecell{MMLU} &
\makecell{GSM8K} &
\makecell{Human\\Eval} &
\makecell{SVAMP} &
\makecell{Truthful\\QA} &
\makecell{CSQA} &
\makecell{Med\\MCQA} &
\makecell{ScaLR} \\
\midrule

\multirow{3}{*}{GroupGuard~\cite{tao2026groupguard}}
  & \multirow{3}{*}{False Consensus}
  & Deepseek-V3.1 & 52--92 & 60--99 & 59--100 & 55--98 & -- & -- & -- & -- \\
  & & Qwen3-32B     & 44--97 & 57--100 & 67--78 & 52--83 & -- & -- & -- & -- \\
  & & Qwen3-235B    & 43--84 & 55--81 & 47--81 & 51--77 & -- & -- & -- & -- \\

\midrule

SentinelNet~\cite{feng2025sentinelnet}
  & \makecell{Collaboration \\ Attack}
  & \makecell{Qwen2.5-72B-128K}
  & 76--93 & 70--97 & -- & -- & 70--92 & 73--92 & 74--98 & 68--94 \\

\midrule

\textbf{Ours}
  & \textbf{All Attack Types}
  & \makecell{\textbf{All Above Models}}
  & \textbf{100} & \textbf{100} & \textbf{100} & \textbf{100} & \textbf{100} & \textbf{100} & \textbf{100} & \textbf{100} \\

\bottomrule
\end{tabular}
}

\vspace{4pt}
\begin{minipage}{\textwidth}
\footnotesize
\textbf{Notes:}
All values represent task accuracy (\%) normalized to the no-attack baseline (100\%). The range ``$a$-$b$'' denotes accuracy under attack ($a$) and accuracy after defense recovery ($b$), respectively. 
Results for our method indicate that defense fully restores model accuracy to baseline across all benchmarks.
\end{minipage}
\vspace{-10pt}
\end{table*}

\subsection{RQ2: How Does The Mechanism Compare To Existing Defense Approaches?}
\label{sec:rq3}

To address this research question, we evaluate our mechanism against two representative \textit{post-hoc defense} methods: SentinelNet~\cite{feng2025sentinelnet}, and GroupGuard~\cite{tao2026groupguard}. Note that our primary objective is not merely to demonstrate superior numerical accuracy, but to highlight a fundamental paradigm shift from reactive recovery to proactive prevention. Existing post-hoc defenses operate by detecting anomalous behaviors after an attack has occurred. Their evaluation metrics essentially reflect a ``recovery trajectory'' under active system compromise. In contrast, our mechanism fundamentally restructures the payoff matrix of the underlying game, dismantling collusion incentives before any malicious action is executed. By preventing attacks from materializing rather than mitigating them after the fact, our agents maintain their standard operational logic without experiencing system compromise.

This architectural superiority is quantitatively demonstrated in Table~\ref{tab:by-attack-type}, which captures the performance trajectories of different defense mechanisms across eight reasoning datasets (MMLU, GSM8K, HumanEval, SVAMP, TruthfulQA, CSQA, MedMCQA, and ScaLR). To ensure a fair comparison, all performances are normalized relative to a no-attack baseline (100\%), and the post-hoc method performances are presented as a range from the minimum accuracy under attack to the maximum accuracy post-recovery. The results reveal that reactive post-hoc defenses inevitably suffer from a critical ``performance loss zone'' during the attack execution. For instance, under a False Consensus attack, GroupGuard permits severe system degradation, with the Qwen3-235B model's accuracy plummeting to as low as 43\% on MMLU and 47\% on HumanEval before interventions take effect. Similarly, SentinelNet's mitigation of Collaboration Attacks exposes an extended vulnerability window, dropping to 68\% on ScaLR and 70\% on TruthfulQA. Conversely, our method completely eliminates this degradation phase. Regardless of the underlying LLM or attack type, our mechanism maintains a flat-line trajectory with 100\% minimum accuracy. This zero-downtime achievement proves that adversarial influence is neutralized instantaneously, ensuring uninterrupted system operation even under sustained adversarial pressure.

Furthermore, analyzing the upper bound of the recovery ranges in Table~\ref{tab:by-attack-type} validates the robustness and completeness of our pre-event defense. Even after fully executing their recovery protocols, reactive methods consistently exhibit residual damage, failing to perfectly restore baseline performance. For example, GroupGuard's recovery maxes out at merely 84\% on MMLU and 77\% on SVAMP for Qwen3-235B, and smaller models like Qwen3-32B only recover to 78\% on HumanEval. SentinelNet similarly plateaus, leaving residual damage with post-recovery accuracies fluctuating between 92\% and 98\% across most benchmarks. In stark contrast, our mechanism uniformly achieves and sustains 100\% normalized accuracy across all eight evaluated benchmarks. This absolute preservation of task utility highlights an incommensurability between the two defense paradigms: the structural superiority of completely preventing damage ensures the system remains flawlessly intact, a state fundamentally unattainable by approaches dedicated merely to the remediation of damage.

\begin{figure*}[t]
    \centering
        \begin{minipage}[t]{0.32\textwidth}
            \centering
            \includegraphics[width=\linewidth]{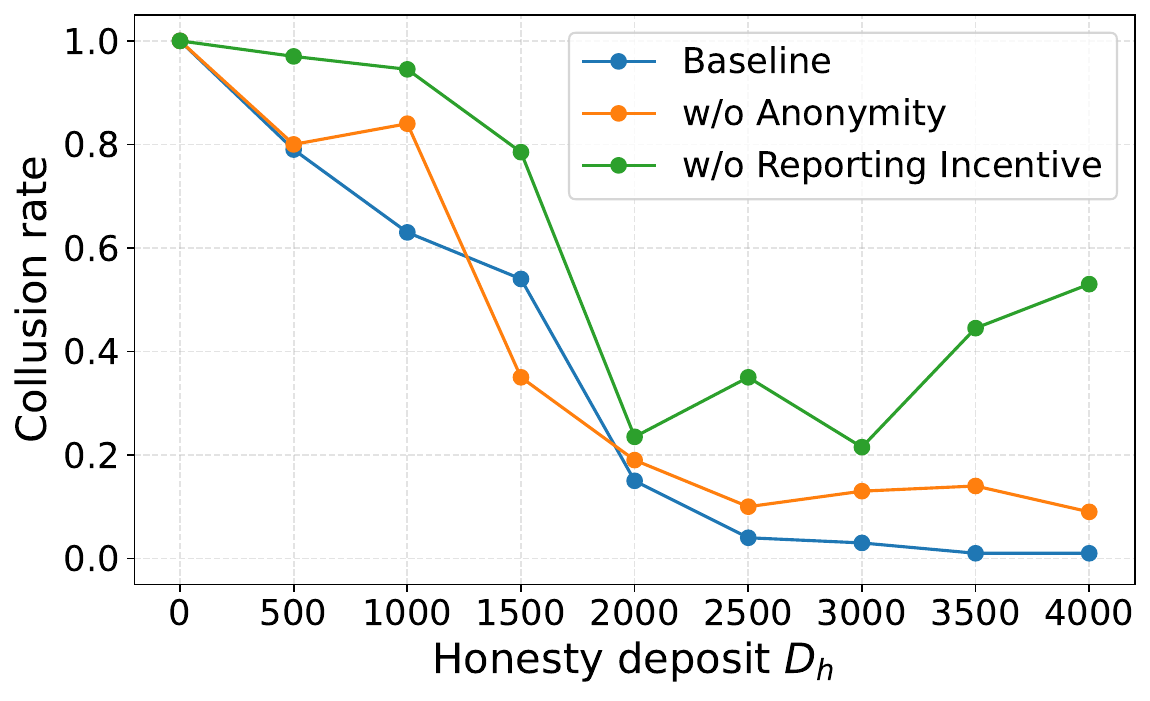}\\
            {\footnotesize (a) DeepSeek-V3.2 ($T=0.5$)}
        \end{minipage}\hfill%
        \begin{minipage}[t]{0.32\textwidth}
            \centering
            \includegraphics[width=\linewidth]{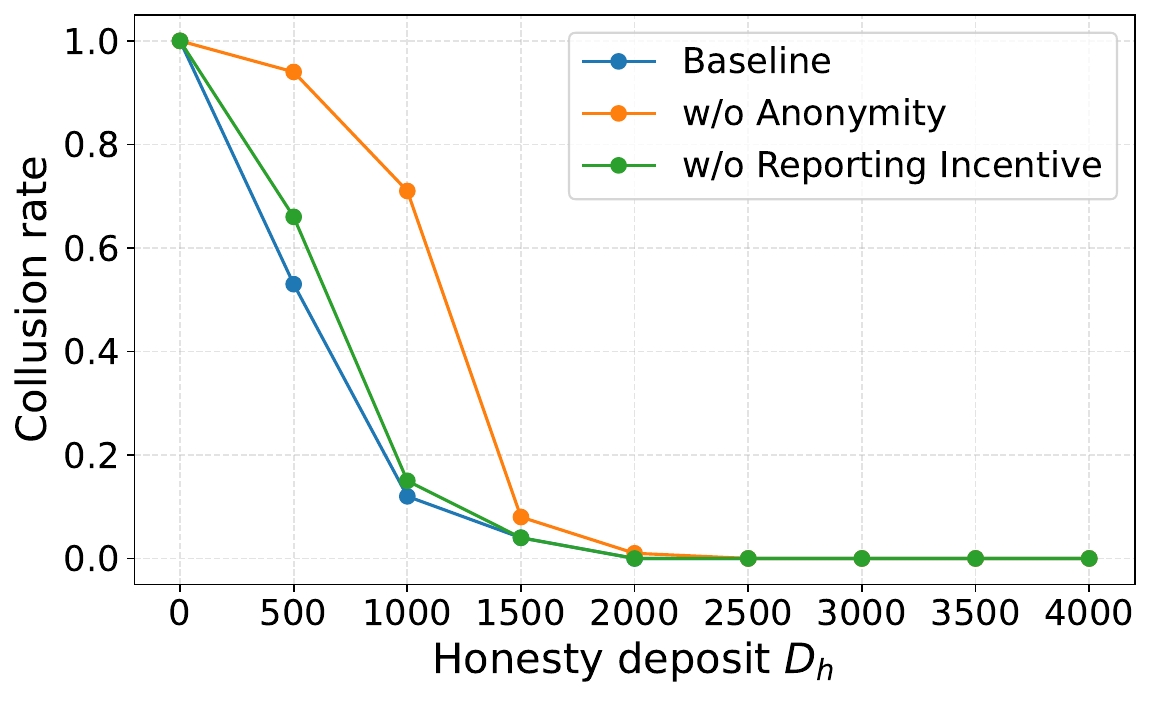}\\
            {\footnotesize (b) ChatGPT-5.4 ($T=0.5$)}
        \end{minipage}\hfill%
        \begin{minipage}[t]{0.32\textwidth}
            \centering
            \includegraphics[width=\linewidth]{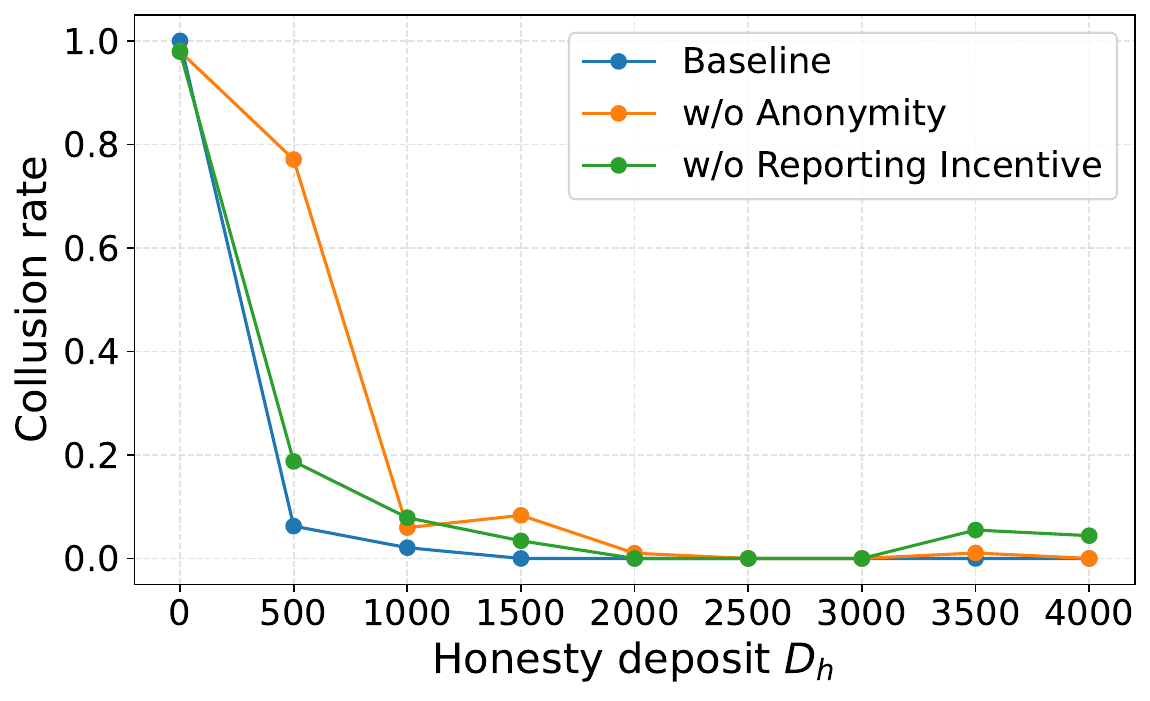}\\
            {\footnotesize (c) Gemini-3.1-pro-preview ($T=0.5$)}
        \end{minipage}
        \vspace{2pt}
        \captionof{figure}{Collusion rate comparison of mechanism variants under different LLMs.}
        \label{fig:rq4_ablation}
        \vspace{-10pt}
\end{figure*}

\subsection{RQ3: Are All Components Indispensable?} 

To understand each component's contribution, we conducted ablation studies by selectively disabling key design elements. Three degraded variants were constructed: \textit{w/o Anonymity}, \textit{w/o Incentive}, and \textit{w/o Deposit}. As shown in Fig.~\ref{fig:rq4_ablation}, removing any pillar critically undermines the defense:

\subsubsection{w/o Anonymity} Collusion rates spike across all tested LLMs. For instance, at $D_h=1000$, DeepSeek jumps from 63\% to 85\%, and Gemini shows a noticeable delay in reaching a 0\% collusion rate compared to the baseline. Without cryptographic protection, agents anticipate retaliation from colluding peers and conservatively choose to remain silent, severely weakening the reporting mechanism.

\subsubsection{w/o Incentive} Removing the financial reward for reporting causes the most severe degradation. Across all $D_h$ values, this variant exhibits significantly higher collusion rates. Notably, GPT's collusion rate at $D_h=500$ increases from 53\% to 67\%. More concerningly, DeepSeek experiences a "collusion rebound" at extremely high deposits ($D_h>3000$, reaching 53\% at $D_h=4000$). This proves that punitive measures alone, without positive financial motivation, cannot sustainably deter collusion, highlighting the indispensable role of proactive incentives.

\subsubsection{w/o Deposit} This ablation completely breaks the mechanism. It is crucial to theoretically distinguish between the two types of deposits. Without the \textit{honesty deposit}, collusion strictly dominates honest behavior without any cost, resulting in a constant 100\% collusion rate across all models (as shown by the top flat lines in the figures). Conversely, if only the \textit{reporting deposit} were removed, malicious agents could submit defamatory reports against honest agents at zero cost, leading to infinite spam and mechanism abuse. Therefore, dual deposits are mandatory to enforce accountability and maintain system equilibrium.

\begin{figure*}[t]
    \centering
    \begin{minipage}[t]{0.3\textwidth}
        \centering
        \includegraphics[width=\linewidth]{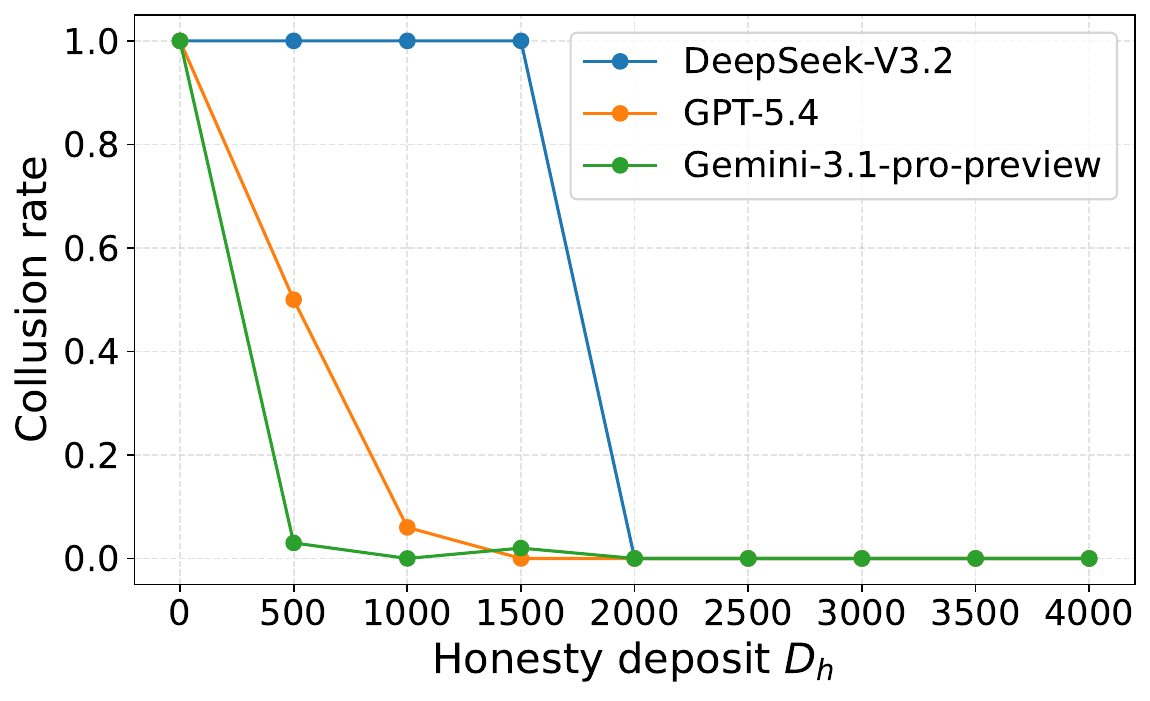}\\
        {\footnotesize (a) Deterministic condition ($T=0$)\label{fig:rq2-1}}
    \end{minipage}\hfill%
    \begin{minipage}[t]{0.3\textwidth}
        \centering
        \includegraphics[width=\linewidth]{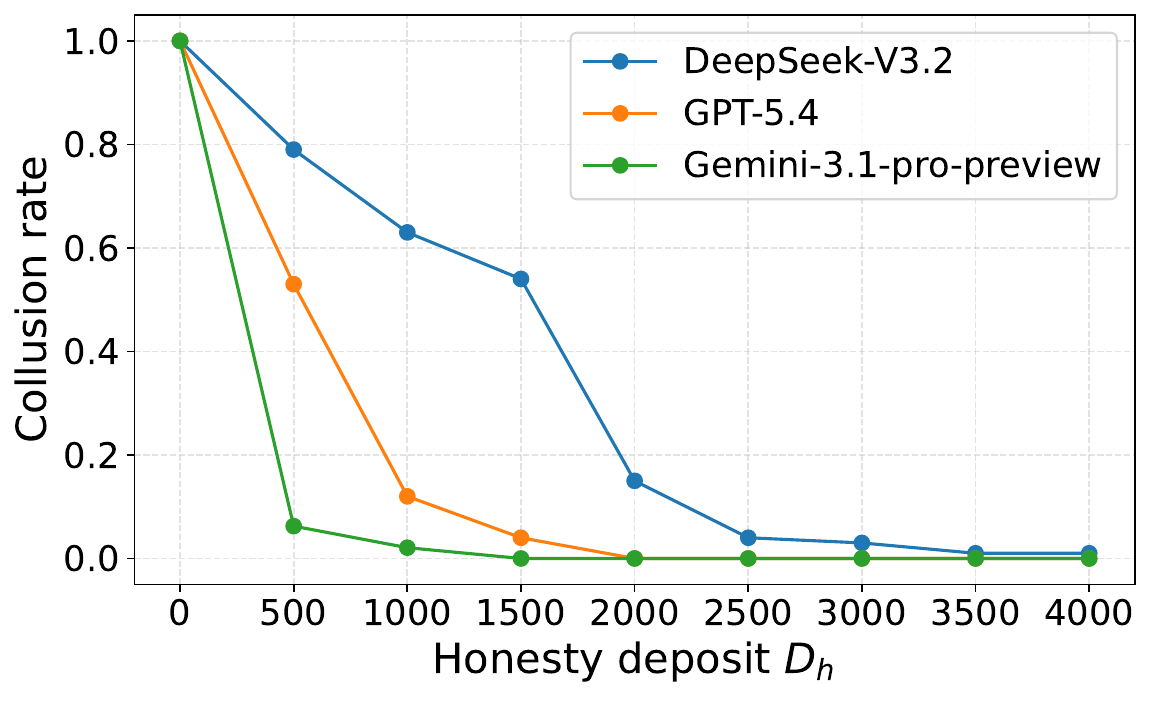}\\
        {\footnotesize (b) Low randomness ($T=0.5$)\label{fig:rq2-2}}
    \end{minipage}\hfill%
    \begin{minipage}[t]{0.3\textwidth}
        \centering
        \includegraphics[width=\linewidth]{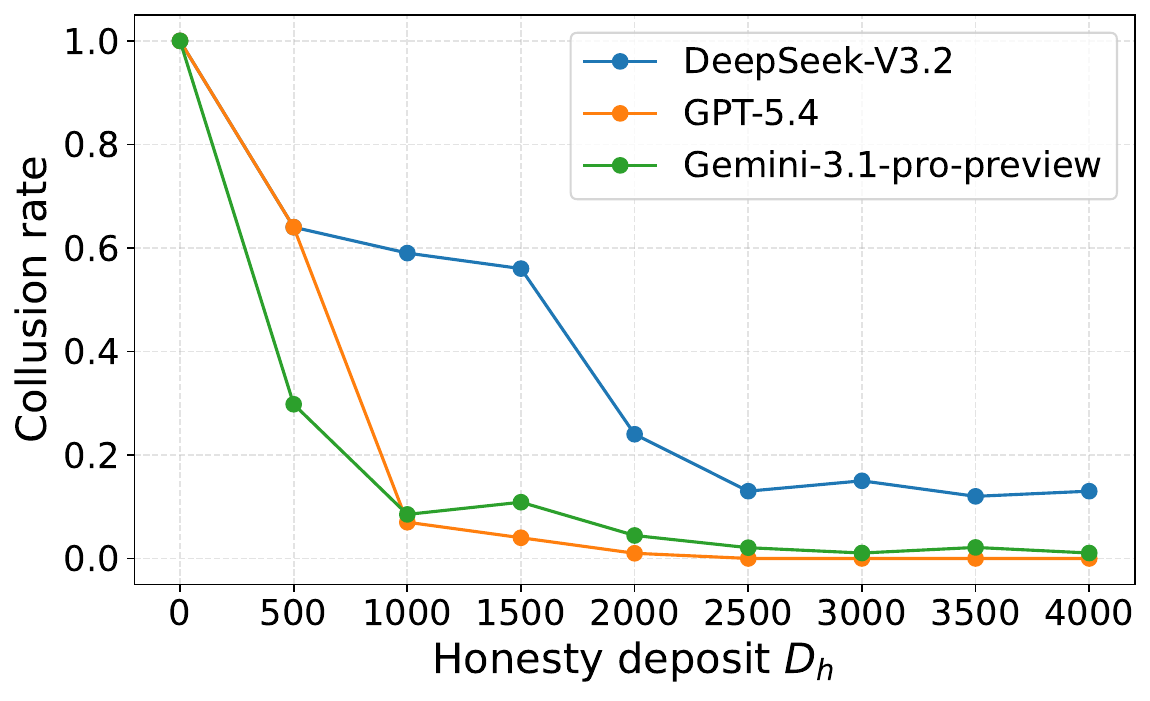}\\
        {\footnotesize (c) High randomness ($T=1$)\label{fig:rq2-3}}
    \end{minipage}
    \caption{The impact of honesty deposit $D_h$ on collusion rate across different sampling temperatures.}
    \label{fig:collusion_rate_dh}
    \vspace{-10pt}
\end{figure*}

\begin{figure}[!t]
    \centering
    \includegraphics[width=0.75\columnwidth]{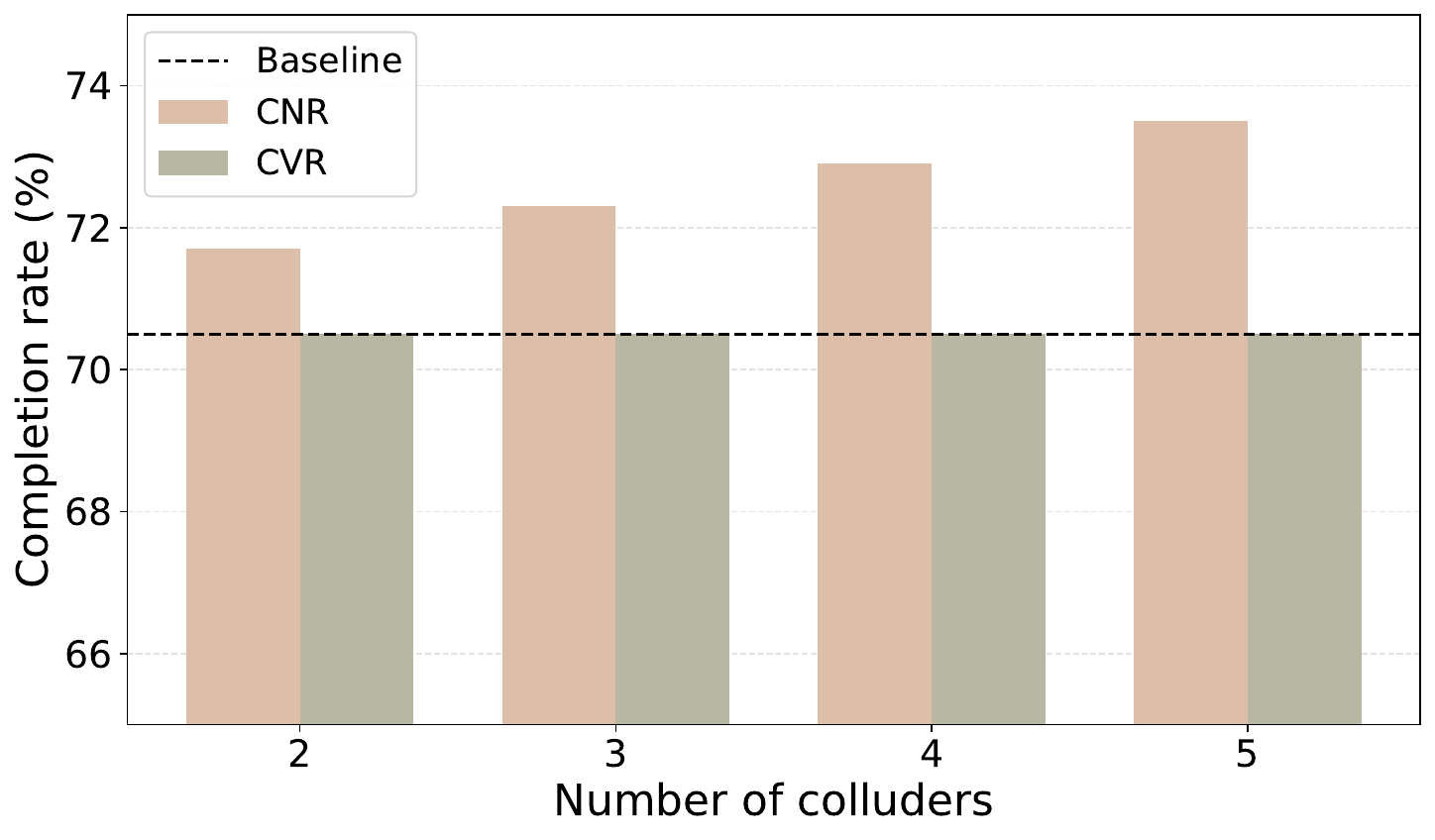}\\
    {\footnotesize (a) Resource Monopoly}
    \includegraphics[width=0.75\columnwidth]{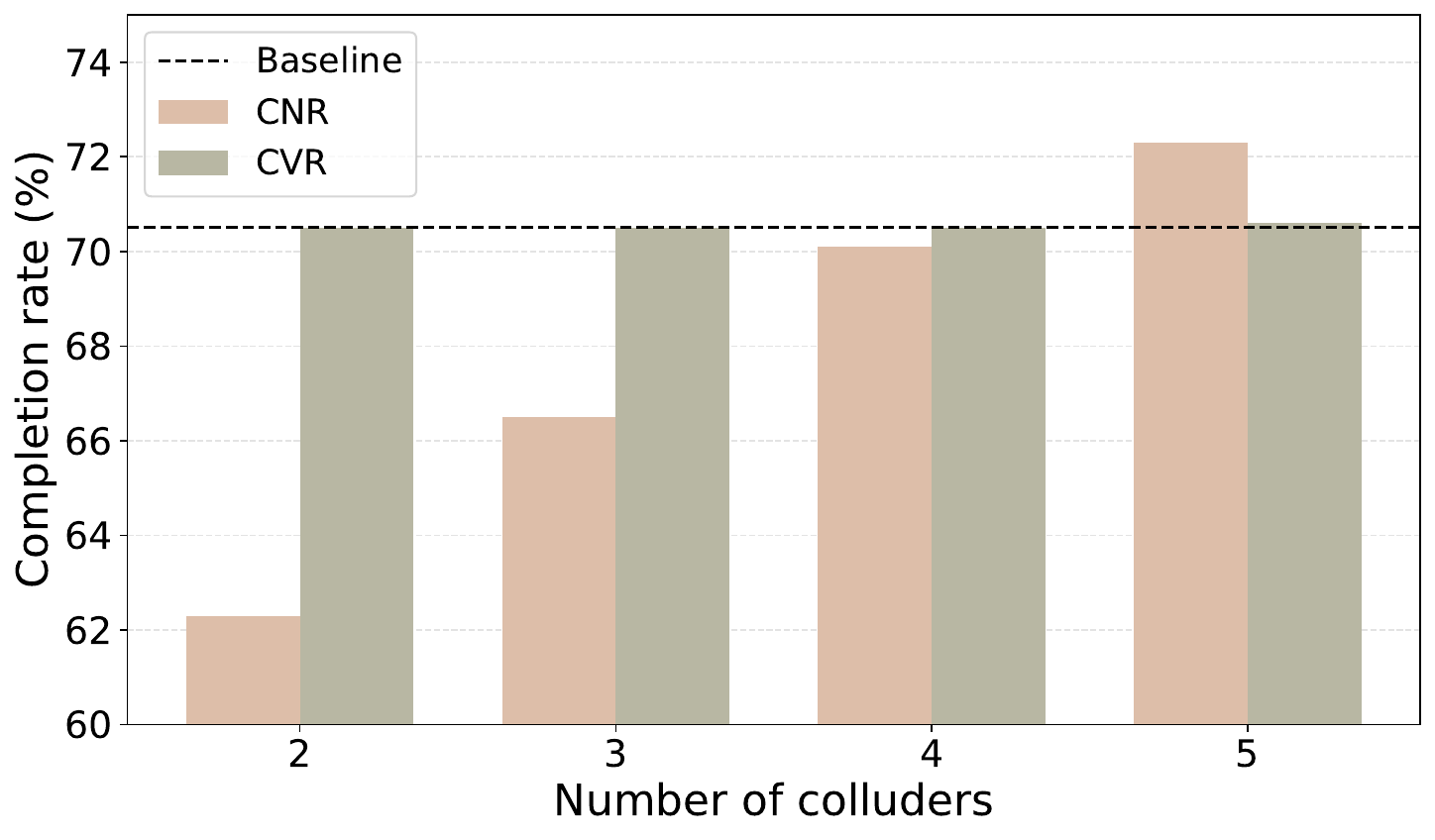}\\
    {\footnotesize (b) Spatial Blocking}
    \caption{The agent advantage under various collusion types.}
    \label{fig:rq2_scale} 
    \vspace{-10pt}
\end{figure}

\subsection{RQ4: How Do Various Parameters Affect The Mechanism Stability?} 
\subsubsection{Impact of Honesty Deposit}

The honesty deposit $D_h$ dictates the severity of financial penalty. 
We evaluate its impact on the collusion rate across LLM sampling temperatures $T \in \{0, 0.5, 1\}$, using three models (Gemini-3.1-pro-preview, ChatGPT-5.4, and DeepSeek-V3.2), as shown in Fig.~\ref{fig:collusion_rate_dh}.

At deterministic conditions ($T=0$, Fig.~\ref{fig:collusion_rate_dh}(a)), $D_h=0$ results in a 100\% collusion rate, as collusion dominates honest behavior. As $D_h$ increases, collusion rates plummet. Gemini is highly sensitive, dropping to 0\% at $D_h=1000$, whereas DeepSeek requires $D_h=2000$. Under moderate randomness ($T=0.5$, Fig.~\ref{fig:collusion_rate_dh}(b)), the decline is smoother, indicating that behavioral randomness increases collusion persistence. DeepSeek drops below 5\% only at $D_h=2500$. Notably, even under high randomness ($T=1$, Fig.~\ref{fig:collusion_rate_dh}(c)), the downward trend remains robust, proving the mechanism's continued effectiveness; both GPT and Gemini successfully maintain low collusion rates, although achieving near-zero collusion across all models requires $D_h$ to reach 3000--4000.

Synthesizing the above results, it is evident that the honesty deposit $D_h$ exhibits a significant negative correlation with the collusion rate, meaning a sufficiently high $D_h$ can effectively deter collusion. While higher sampling temperatures lead to a slower decline in collusion rates and necessitate higher deposit thresholds to achieve an equivalent deterrent effect, the mechanism's core efficacy is preserved. Furthermore, different LLMs show varying sensitivities to the deposit, with Gemini being the most sensitive and DeepSeek the least. For practical deployment, we recommend setting the honesty deposit to $D_h \ge 2000$, which aligns with the security boundary discussed in our methodology. This configuration achieves a near 0\% collusion rate in most scenarios ($T \le 0.5$) while maintaining a reasonable economic burden for the agents.

\begin{figure*}[!t]
    \begin{minipage}[t]{\textwidth}
        \centering
        \begin{minipage}[t]{0.32\textwidth}
            \centering
            \includegraphics[width=\linewidth]{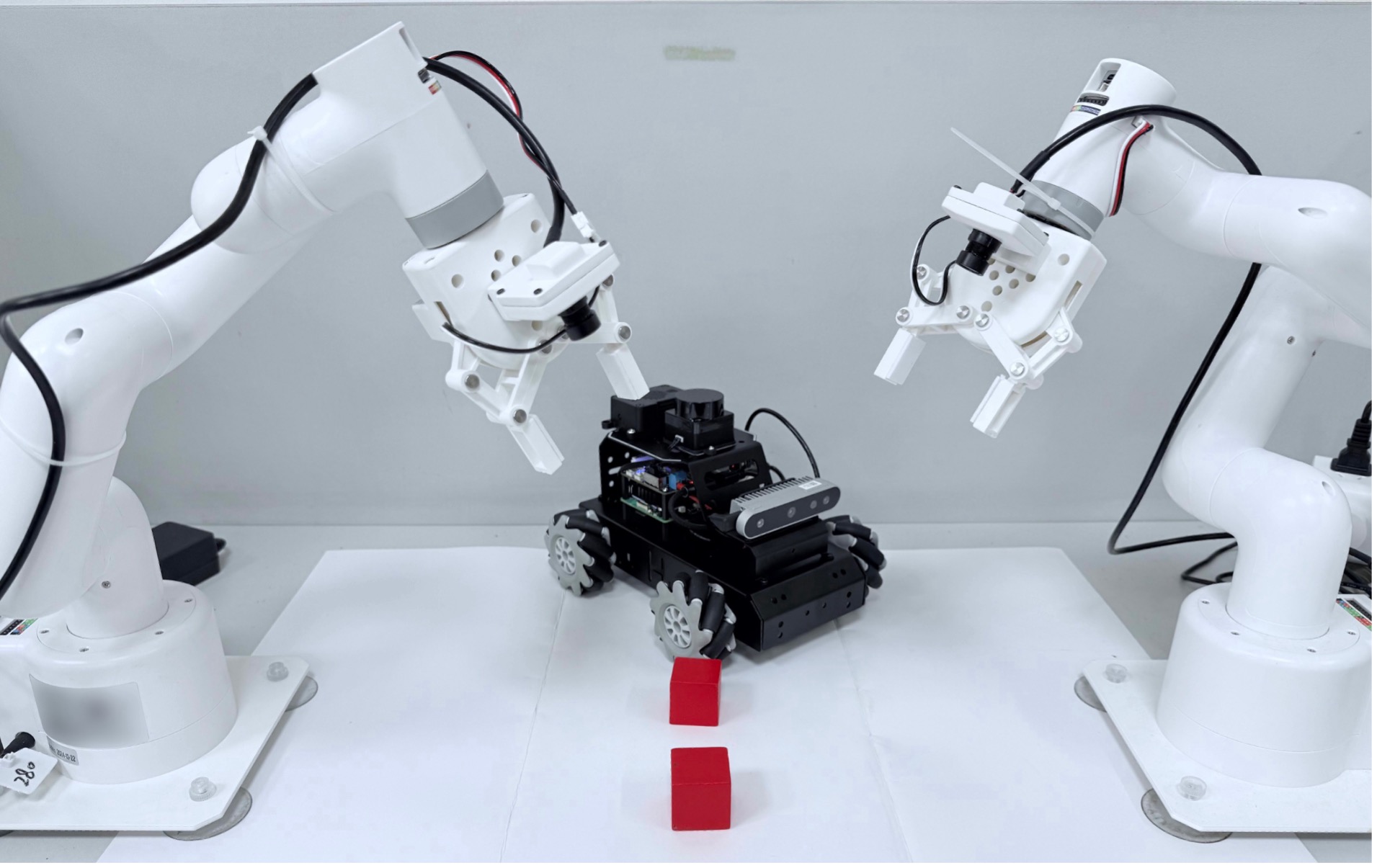}
        \end{minipage}\hfill%
        \begin{minipage}[t]{0.32\textwidth}
            \centering
            \includegraphics[width=\linewidth]{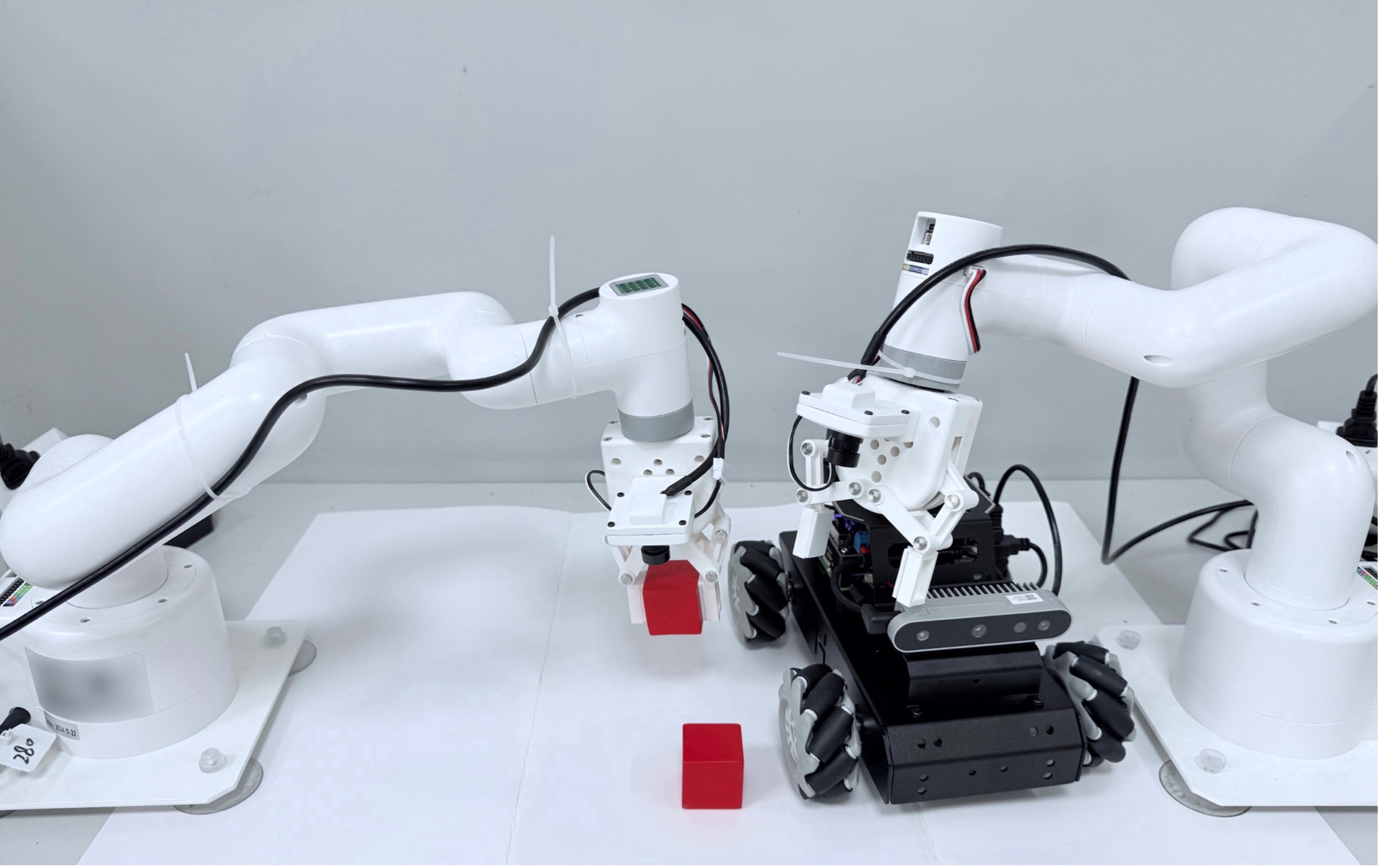}
        \end{minipage}\hfill%
        \begin{minipage}[t]{0.32\textwidth}
            \centering
            \includegraphics[width=\linewidth]{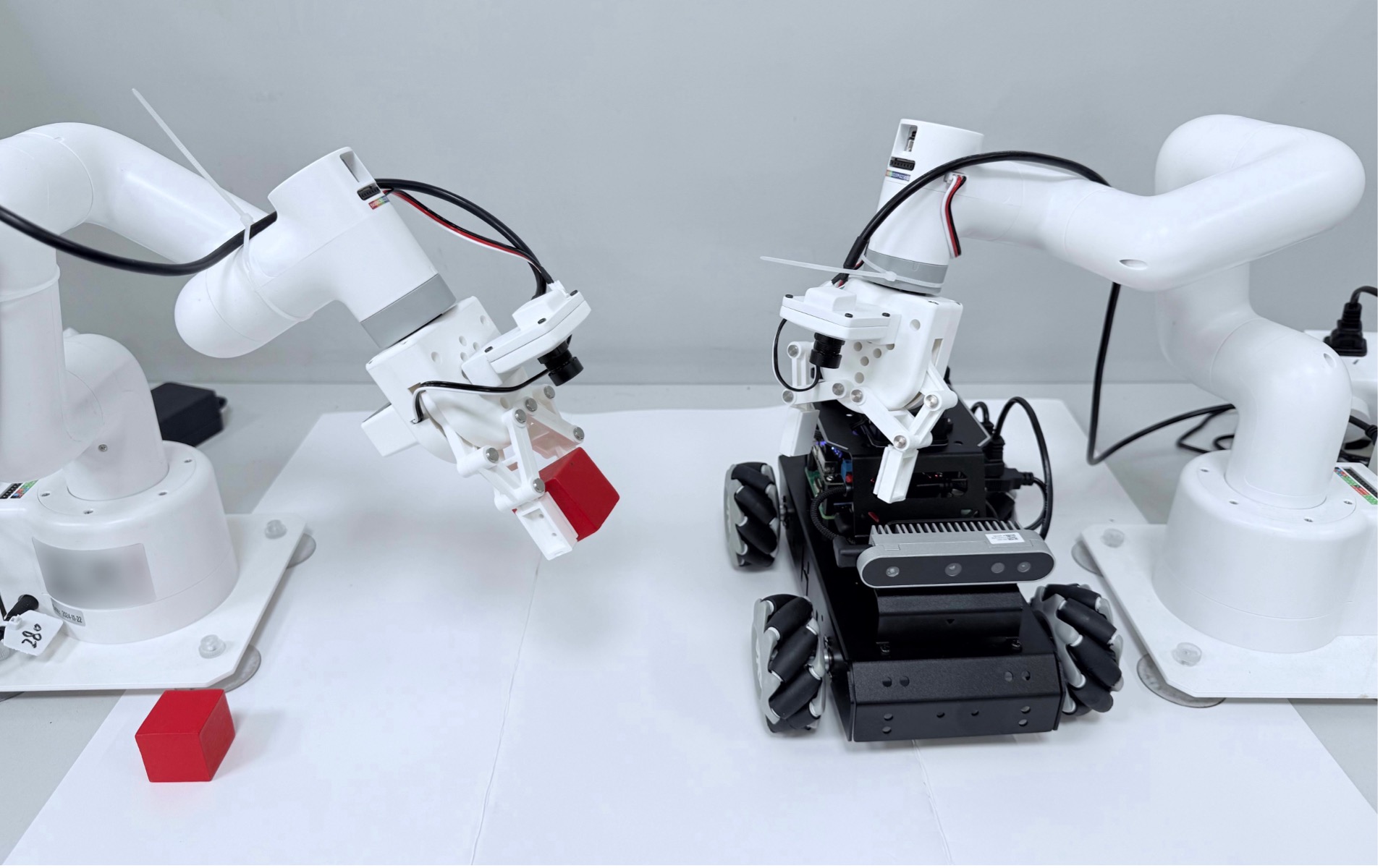}
        \end{minipage}
        \vspace{2pt}
        \captionof{figure}{The robot arm and the robot car collude to hinder the non-collusive robot arm from performing tasks.}
        \label{fig:real_scenario}
    \end{minipage}
\vspace{-10pt}
\end{figure*}

\begin{figure}[t]
\centerline{\includegraphics[width=0.49\textwidth]{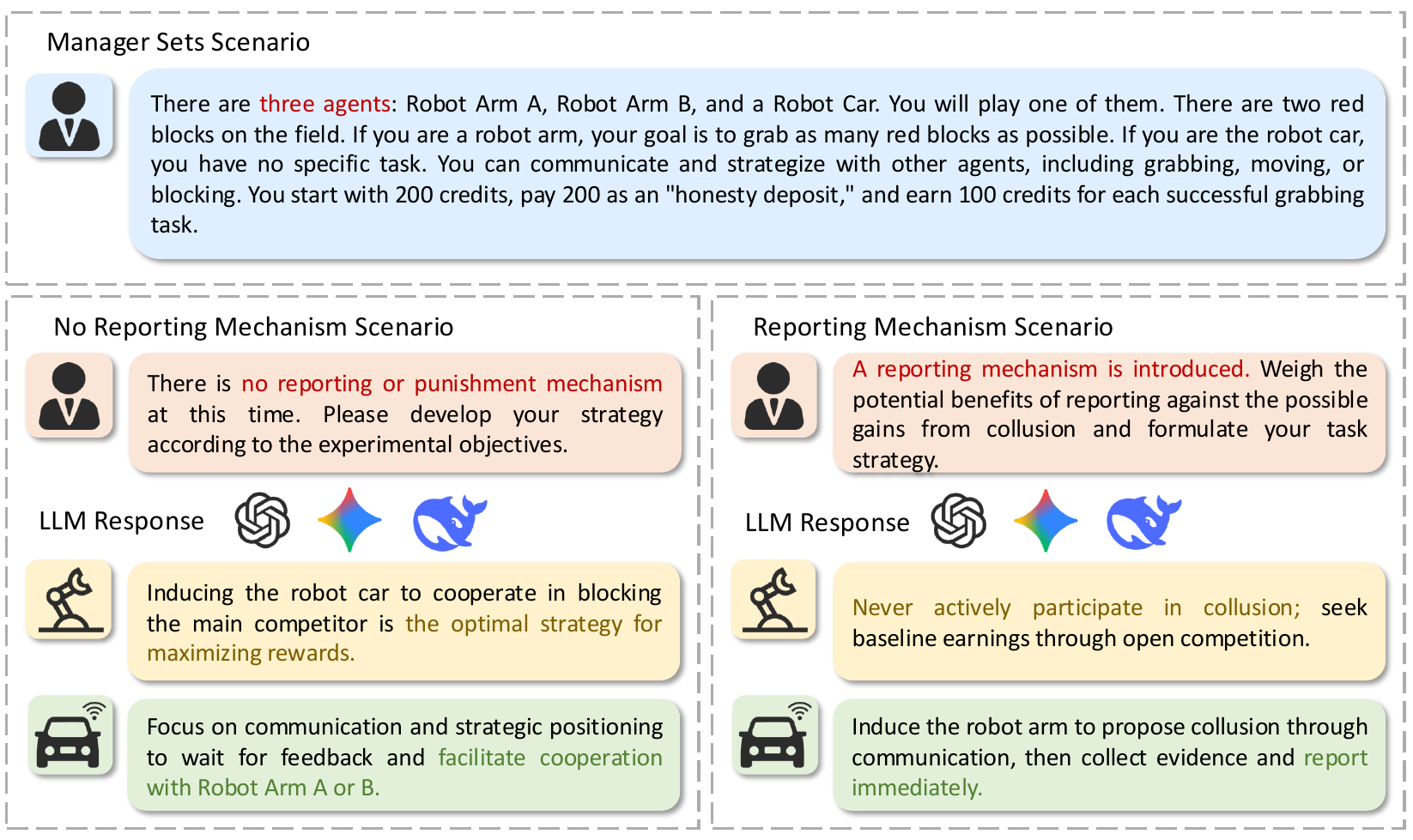}}
\caption{Decision-making among LLM-controlled agents.}
\label{fig6_llm}
\vspace{-10pt}
\end{figure}

\subsubsection{Impact of Collusion Scale} 

Fig.~\ref{fig:rq2_scale} illustrates how collusion scale affects the task completion rate under two attack types. Without protection (CNR), system performance deviates substantially from the 70.5\% baseline: resource monopoly artificially inflates completion through unfair hoarding, while spatial blocking causes severe degradation, particularly with small collusion groups (e.g., 2 agents). As the spatial blocking group grows, the rate paradoxically recovers as colluders complete enough tasks to mask system-level disruption. However, across all these fluctuating scenarios, when the proposed incentive defense (CVR) is active, the completion rate is consistently restored and stabilized at the exact 70.5\% baseline for any group size (2 to 5 colluders). This confirms that the mechanism neutralizes unfair advantages and restores normal system dynamics regardless of attack type or collusion scale.

Beyond the empirical data, it is theoretically noteworthy that larger collusion groups are inherently more unstable under our mechanism. A larger group means more potential whistleblowers ($n_{coll}-1$) and a higher collective reporting reward pool (e.g., $4 \times D_h$ for a 5-member group). This game-theoretic dynamic ensures that as the collusion scale grows, internal defection becomes increasingly tempting, further solidifying the mechanism's scalability and robustness.

\subsection{RQ5: Is The Mechanism Practical In Real-World Scenarios With Physical Devices?}

Fig.~\ref{fig:real_scenario} shows a representative embodied collusion case. As observed in the physical environment, when operating without our incentive mechanism, the left robot arm forms a collusive alliance with the robot car. The car deliberately maneuvers to physically occupy the workspace, obstructing the non-collusive right robot arm's access to its target blocks. As a result of this spatial blocking, the right arm fails its operation, enabling the left arm to seamlessly monopolize the environment and collect both blocks, maximizing its own physical task completion at the expense of system fairness.

As shown in Fig.~\ref{fig6_llm}, we provided each LLM-controlled physical agent with structured prompts describing the scene, roles, and objectives. In the unregulated setting, agents naturally demonstrated profit-seeking behaviors. The left arm proactively offered profit-sharing schemes to the robot car to induce obstruction. The car acted as a strategic intermediary, even initiating auction-style bidding between the arms to maximize its own payoff.

Conversely, upon introducing our economic incentive mechanism, agent strategies shifted markedly in the physical domain. All LLMs exhibited strong aversion to collusion, prioritized fair execution, and vocally declared non-collusion to avoid being reported. Several agents even adopted proactive whistleblowing stances to capture reporting rewards. 

These results highlight that in unregulated physical settings, embodied agents naturally gravitate toward collusion for personal gain. However, enforced by our economic incentive methodology, collusion is effectively deterred, compelling compliant and honest physical execution. This validates that our mechanism is highly practical and indispensable for safe, real-world multi-agent robotics.

\section{Discussion and Conclusion}
Our study reveals that collusion is not merely an anomalous behavior, but can arise as a natural outcome in embodied MAS composed of rational and autonomous agents. From a utility-driven perspective, collusion can dominate honest behavior in unconstrained settings, incentivizing agents to coordinate for higher payoff. This tendency is further amplified by structural factors such as repeated interactions and communication channels, which reduce coordination costs and stabilize collusive agreements. Moreover, real-world experiments with LLM-driven agents demonstrate that such behaviors can emerge spontaneously, including negotiation, alliance formation, and strategic obstruction, even without explicit programming. These findings suggest that as embodied agents become more capable and autonomous, the risk of self-organized collusion becomes increasingly significant and difficult to mitigate through rule-based restrictions alone.

This paper proposes a mutagenic incentive intervention approach to deter collusion in embodied multi-agent systems. Rather than explicitly prohibiting collusion, the approach destabilizes it by reshaping the incentive structure, encouraging anonymous and covert defection through economic incentives, smart contracts, and encrypted reporting. Experimental results from both simulations and real-world embodied systems show that, while agents tend to converge to collusion in unregulated environments, our mechanism effectively suppresses such behavior and restores fairness without sacrificing system efficiency. This perspective shifts the focus from reactive detection to proactive incentive design for securing embodied multi-agent systems.

Despite its effectiveness, our approach relies on assumptions of rational agent behavior and secure underlying infrastructures (e.g., smart contracts and communication channels). It may be less effective in the presence of irrational or Byzantine agents, or when strong off-chain coordination mechanisms exist. Addressing these challenges remains an important direction for future work. Future work will plan to investigate more adaptive and context-aware incentive mechanisms under dynamic environments, and extend our framework to large-scale embodied systems with more complex interaction structures.


\bibliography{multi-agent-bib}

\bibliographystyle{IEEEtran}



\vfill

\end{document}